\documentclass[12pt, epsfig]{article}

\usepackage[hmargin=2.3cm,vmargin=2.7cm]{geometry}
\usepackage{times}
\usepackage{graphicx}

\begin{document}
\title{Gravitational backreaction in cosmological spacetimes  }
\author{Charis Anastopoulos\footnote{anastop@physics.upatras.gr}\\
{\small Department of Physics, University of Patras, 26500 Patras,
Greece}} \maketitle

\begin{abstract}
We develop a new formalism for the treatment of gravitational
backreaction in the cosmological setting. The approach is inspired
by projective techniques in non-equilibrium statistical mechanics.
We employ group-averaging with respect to the action of the isotropy
group of homogeneous and isotropic spacetimes (rather than spatial
averaging), in order to define effective FRW variables for a generic
spacetime. Using the Hamiltonian formalism for gravitating perfect
fluids, we obtain a set of  equations for the evolution of the
effective variables; these equations incorporate the effects of
backreaction by the inhomogeneities. Specializing to dust-filled
spacetimes, we find regimes that lead to a closed set of
backreaction equations, which we solve for small inhomogeneities. We
then study the case of large inhomogeneities in relation to the
proposal that backreaction can lead to accelerated expansion. In
particular, we identify regions of the gravitational state space
that correspond
 to effective cosmic acceleration. Necessary conditions are (i) a strong expansion
 of the congruences corresponding to comoving observers, and (ii) a
 large negative value of a dissipation variable that appears
 in the effective equations (i.e, an effective "anti-dissipation").
\end{abstract}

\section{Introduction}

\subsection{Preamble}
The fundamental postulate of modern cosmology is the assumption of a
homogeneous and isotropic universe. The spacetime must then possess
a six-dimensional group of spacelike isometries, i.e., it must be of
the   Friedmann-Robertsn-Walker (FRW) type. However, isotropy and
homogeneity refer to a coarse-grained level of description: there is
significant inhomogeneity at short length-scales.

Since homogeneity is approximate, one may inquire how
inhomogeneities affect the evolution  of the FRW variables. This
question is of foundational interest, as it touches upon the domain
of validity of the fundamental assumption of modern cosmology
\cite{ellis1, back}. Moreover, both in early universe cosmology
(inflation in particular) and in present-epoch cosmology,
gravitational backreaction effects may play a significant role in
the evolution of the universe. In particular, it has been proposed
that the backreaction of spatial inhomogeneities may be responsible
for the apparent cosmic acceleration, so that one would not have to
invoke the existence of dark energy \cite{ltb, de, swiss}--see also
the reviews \cite{celer, buchert} and critique \cite{WI}.

In this paper we develop a systematic approach for the study of
gravitational backreaction with an emphasis on the cosmological
context. Our starting point is the observation that the issue of
gravitational backreaction has many analogues in non-equilibrium
statistical mechanics. The description of a spacetime by a single
quantity (the scale factor),  obtained from the "averaging" of a
generic metric, resembles the description of many-body systems
through coarse-grained bulk variables (the mean field approximation
in particular) \cite{DGM, Zwan, Bal, CaHu}. A key problem in the
study of gravitational backreaction is the consistency of the
approximation scheme. Problems of this type are specifically
addressed by the techniques of non-equilibrium statistical
mechanics. A transfer of ideas from this field of research for the
study of gravitational perturbations  could prove highly fruitful.

A second problem in treatments of gravitational backreaction arises
from the issue of gauge invariance. A definition of effective FRW
variables involves spatial averaging of the inhomogeneities, and
such averaging can be implemented covariantly only for spatial
scalars. For this reason,  studies of backreaction  often restrict
to averaging  the Hamiltonian constraint, which is a spatial scalar;
the averaging cannot be implemented at the level of the
tensor-field-valued equations of motion. Moreover, even if one
restricts to scalars, the results depend on the choice of foliation,
i.e., on the family of spatial surfaces upon which one integrates.
Lack of gauge invariance may also pose a problem in another stage,
namely,  in the implementation of the dynamics. If the dynamics of
gravitational perturbations are described through gauge-fixing, then
there is the danger that that the study of back-reaction will lead
to results that depend on the chosen gauge \cite{WI}.

The formalism we develop in this paper is  gauge-invariant. The main
difference from previous approaches is that  we employ {\em group
averaging} over the isometry group of the FRW spacetime, rather than
spatial averaging. Group averaging is  defined covariantly for any
tensor field, and it reduces to spatial averaging for scalar
quantities. As a matter of fact, the properties of the
group-averaging calculus allow all group averages that appear in
this paper to be reduced to spatial averages of scalar quantities.

In order to avoid the problems related to gauge-fixing, we work
within the Hamiltonian formalism for general relativity. In this
paper, we also restrict the matter content to perfect fluids.In this
case, the solution of the constraints can be implemented in a fully
geometrical way (i.e., without gauge fixing) at the level of the
Lagrangian. The reduction entails a solution of the diffeomorphism
constraints: this corresponds in an implicit "selection" of a class
of foliations tied to the perfect fluid. The only "gauge choice"
that remains is that of a time variable. The formalism allows the
derivation of backreaction equations for any such choice, as long as
it can be made consistently over the system's state space;  we
choose time as measured by observers comoving with the fluid.

\subsection{Backreaction and non-equilibrium statistical mechanics}

A common approach for the study of backreaction effects in a
cosmological spacetime employs a perturbation expansion around the
FRW solution. This involves solving the linearized Einstein
equations around the classical FRW solution; the perturbations are
then employed for the construction of an effective stress-energy
tensor, which, when inserted into the Einstein equations, provides
the corrections of the FRW evolution. In general, this procedure
suffers from consistency problems. In particular, the construction
of the effective stress-energy tensor is gauge-dependent--hence the
resulting backreaction equations are also gauge-dependent \cite{WI}.

Moreover, when large perturbations are taken into account, the
accuracy of such an approximation scheme  degenerates rapidly with
time: the solution of the equations of motion with backreaction
diverges cumulatively from the FRW solution. This means that the
perturbations around the FRW equations capture less and less of the
physics of the system as time increases. Outside the gravitational
context, such treatments  are known to misrepresent
backreaction-induced effects such as dissipation and diffusion.

The consistent treatment of backreaction is a major ingredient in
most techniques developed in the field of non-equilibrium
statistical mechanics. The methodology of such treatments varies
according to the system under consideration. However,  all
treatments follow a common pattern, which is abstractly and
compactly described in the language of the so-called projection
formalism \cite{Zwan, proj}. This pattern can be described as a
sequence of three steps.

The first step is the specification of the {\em level of
description}, namely, of  a set of variables that provide a
 coarse-grained description of the  system under
consideration. For example, in quantum Brownian motion, one studies
a selected particle interacting with a heat bath of harmonic
oscillators (environment). The level of description corresponds to
the degrees of freedom of the selected particle. In Boltzmann's
treatment of the rare gas, the system is a collection of weakly
interacting particles, and
 the level of description
is defined by a probability density on the phase space of a {\em
single} particle. In general, the level of description corresponds
to a subspace of the space of functions $F(\Gamma)$ on the system's
state space $\Gamma$.
 It is
represented by a projective map $P$ on  $F(\Gamma)$. We shall call
the variables that lie within the range of $P$ {\em relevant
variables} and ones that lie on the range of $1 - P$ {\em
non-relevant} variables. For the cosmological perturbations
considered here, the relevant variables correspond to homogeneous
and isotropic field configurations.

The second step involves a splitting of the dynamical evolution into
components in accordance with the chosen level of description. Let
$L_t$ be the evolution operator on the space of states (the
propagator of the Liouville equation in a Hamiltonian system).

\begin{itemize}
\item $PL_tP$ describes the self-evolution of the   relevant
variables.
\item  $(1-P) L_t (1 - P)$ describes the self-evolution of the
 non-relevant variables.
\item $(1-P)L_t P$ describes the coupling between relevant and
non-relevant variables.
\end{itemize}

The splitting above allows for the derivation of a set of evolution
equation for the relevant variables: this contains the evolution
terms $PL_tP$ for the relevant variables and  backreaction terms
that arise from the coupling $(1-P)L_t P$ of relevant to
non-relevant variables. The backreaction terms
 depend on the state of the non-relevant variables and on their
 self-evolution in terms of $(1-P) L_t (1 - P)$. In general, the set of evolution equations
for the relevant variables
 is  not autonomous, or not-closed, and bears an explicit dependence
 on
 the initial state of the non-relevant variables.

 The third step is the derivation of a {\em closed} set
of equations (e.g., a Fokker-Planck-type equation for quantum
Brownian motion or Boltzmann's equation for the rare gas). To this
end,  one introduces additional assumptions about the state and
evolution of the non-relevant variables. For example, one may assume
that the state of the non-relevant variables is not significantly
affected by the evolution of the relevant variables (Born
approximation), or that the memory effects in the evolution are
negligible (Markov approximation), or that the non-relevant
variables are "fast" in relation to the relevant ones, or that the
correlations between relevant variables are insignificant (e.g., in
the truncation of the Bogolubov-Born-Kirkwood-Green-Yvon  hierarchy
in kinetic theory), and so on. Such assumptions are necessary for
the closure of the set of evolution equations and they often involve
the introduction of semi-phenomenological parameters that describe
the properties of the irrelevant variables.

To summarize, the general procedure for the consistent construction
of backreaction equations involves:

\begin{enumerate}
 \item a specification of the level of description.
\item  a splitting of the dynamics and the construction of equations
for the relevant variables
\item  additional assumptions about the state of the irrelevant
variables that allow for the closure of the system of effective
equations.
\end{enumerate}

\subsection{Our approach}

In this paper, we apply the reasoning above to the treatment of the
backreaction from cosmological inhomogeneities. The rationale is
that the FRW variables provide a coarse-grained level of description
for a cosmological spacetime, which is, in a sense, analogous to the
description of statistical systems in terms of mean-field theory.
The backreaction of inhomogeneities is then conceptually similar to
the incorporation of the effects of the second- and higher-order
correlation functions into the mean-field evolution.\footnote{We
also note that other techniques from non-equilibrium statistical
mechanics (mainly functional methods for quantum fields) have been
employed in the context of early Universe cosmology--see \cite{CaHu}
and references therein. }

Our primary aim in this paper is to set the basis of a general
procedure for the treatment of inhomogeneities. The projective
formalism can, in principle, be applied to spacetimes with arbitrary
matter content. Specializing to the case of a gravitating perfect
fluid allows us to solve the diffeomorphism constraints in a
gauge-invariant way. We then analyze in detail the special case of
backreaction in a dust-filled spacetime.

We do not assume a specific form for the "true" spacetime metric,
which is to be approximated by an FRW spacetime. Rather, we
construct backreaction equations in terms of a small number of
variables that are defined for {\em any} spatial geometry, i.e., the
backreaction variables are functions defined over  the full
gravitational state space. Hence, the formalism can, in principle,
accommodate any choice for the "true" spacetime metric. Each choice
corresponds to different a region of the gravitational phase space,
in which the backreaction variables take different values and may
generate {\em qualitatively different} evolutions.

 The method  developed here implements steps 1 and
2 described in Sec. 1.2, in a gauge-invariant way. The level of
description is the subspace of the gravitational phase space with
initial data invariant under the action of the six-dimensional
isometry group of the FRW spacetimes, and the projector $P$ is
constructed by the group averaging of observables. Evolution
equations for the relevant variables are obtained through the
Hamiltonian formalism.

The third step involves assumptions about the nature of cosmological
perturbations, i.e., it requires the specification of a region on
the gravitational state space. This ought to be an observational,
rather than a theoretical input, because the intuitions from
non-equilibrium statistical mechanics are not directly relevant to
the cosmological context. For small perturbations, we find generic
regimes that lead to a closed set of backreaction equations, which
can solve explicitly.

For large perturbations, additional variables appear in the
effective evolution equations. It is, therefore, more difficult to
obtain a closed set of backreaction equations. For this reason, in
this paper, we only study the {\em kinematics} of backreaction in
the large perturbations regime. In particular, we examine whether
large backreaction effects can lead to an effective accelerated
expansion. We find that there exists a plausible regime in the
gravitational state space that  manifests effective cosmic
acceleration. This regime corresponds to initial data (present era)
in the gravitational state space that satisfy a small number of
conditions: these conditions are rather restrictive but they are
generic, in the sense that they do not require a "fine-tuning" of
parameters. We find that acceleration necessitates a   strong
expansion
 of the congruences corresponding to comoving observers (i.e., an "intrinsic" expansion of the
 inhomogeneous regions, in addition
 to the Hubble expansion), and  a
 large negative value of a "dissipation" variable that appears
 in the effective equations.

However, the physical relevance of this regime remains an open
issue: it is necessary to demonstrate that the evolution of  initial
data in this regime correspond to a cosmological histories
compatible with observations. Such a demonstration requires a full
dynamical treatment for large perturbations, and it will be taken up
in another work.

It is important to emphasize that in our approach the relevant
variables are determined through integration over the spacetime
group of isometries. In group-averaging the whole set of points of a
Cauchy surface of a cosmological spacetime is involved; specific
subsets cannot be isolated in a coordinate-independent way. For this
reason, the method, in its present form, cannot provide a definition
of an "average geometry" for a generic spatial region. The scope is,
therefore, different from that of the approaches reviewed in
\cite{buchert}. In particular, we do not aim to provide an answer to
questions such as "how does the Universe look at different scales?"
\cite{ellis1}.

\subsection{Structure of the paper}

In Sec. 2, we introduce the notion of group averaging in the context
of general relativity, we study its properties and develop the
calculational tools needed in the remaining of the paper. In Sec.3
we present the Hamiltonian formalism of perfect fluids (in the
Lagrangian rather than the Eulerian picture), following the
treatment of Ref. \cite{KSG}; we show that the diffeomorphism
constraints can be implemented without gauge-fixing.

In Sec. 4, we define the relevant variables for the treatment of
backreaction, we elaborate on the Hamiltonian equations of motion,
and we construct the backreaction equations. In Sec. 5, we study the
evolution of the backreaction parameters in the regime of
 small perturbations. We specialize to dust-filled spacetimes and
we identify regimes that lead to closed sets of backreaction
equations. Large perturbations are taken up in Sec. 6: we find the
corresponding backreaction equations, when spatial curvature effects
are negligible, and we show that there are regimes that correspond
to accelerated expansion. In the final section, we summarize our
results, and briefly discuss possible extensions.

\section{Group averaging}

As we explained in the introduction, our aim is to incorporate the
backreaction of the inhomogeneities into the evolution of
homogeneous and isotropic FRW metrics. To this end, we must
construct a map that takes generic inhomogeneous tensorial variables
(for example, a Riemannian metric) into variables of the same type
compatible with homogeneity and isotropy. A naive way to proceed
would be to integrate the tensor field over a Cauchy surface and
divide by the volume. However, coordinate-invariant integration can
only be defined for scalar fields: spatial integration of a generic
tensor field is not an invariant procedure. In particular, it is not
possible to define an average metric through this method.

The method we develop here is based on group-averaging rather than
spatial averaging. The method is suggested by  the fact that the
{\em defining} feature of the FRW metric is the existence of a
six-dimensional group of isometries.

\subsection{The basic construction}
  Let $G$ be a compact Lie group
acting on a compact three-manifold $\Sigma$. This means that there
exists a smooth map $f: G \rightarrow Diff(\Sigma)$, such that
$f_{g_1} \circ f_{g_2} = f_{g_1 g_2}$, for all $g_1, g_2 \in G$.

\medskip

For any tensor field $A^{i_1 \ldots i_n}{}_{j_1 \ldots j_m}$ on
$\Sigma$, we define the group-averaged tensor field $\langle A^{i_1
\ldots i_n}{}_{j_1 \ldots j_m} \rangle$ as
\begin{eqnarray}
\langle A^{i_1 \ldots i_n}{}_{j_1 \ldots j_m} \rangle(x) = \int d
\mu[g] (f^*_gA)^{i_1 \ldots i_n}{}_{j_1 \ldots j_m}(x), \label{aver}
\end{eqnarray}
where $d \mu[g]$ is the left-invariant Haar measure on $G$
(normalized to unity).

The invariance of $d \mu[g]$ implies that
\begin{eqnarray}
[f^*_g\langle A^{i_1 \ldots i_n}{}_{j_1 \ldots j_m} \rangle](x) =
\langle A^{i_1 \ldots i_n}{}_{j_1 \ldots j_m} \rangle(x).
\end{eqnarray}
It follows that there exist vector fields $X_A$ on $\Sigma$ that
correspond to elements $A$ of the Lie algebra of G, such that
\begin{eqnarray}
{\cal L}_{X_A} \langle A^{i_1 \ldots i_n}{}_{j_1 \ldots j_m} \rangle
= 0.
\end{eqnarray}

Let $\Gamma$ be the state space of Hamiltonian initial data
$(h_{ij}, \pi^{ij})$ for the gravitational field on a spacetime with
Cauchy surfaces of topology $\Sigma$, where $h_{ij}$ is a Riemannian
metric on $\Sigma$ and $\pi^{ij}$ the conjugate momentum (a tensor
density).  $\Gamma = T^*Riem(\Sigma)$, where $Riem(\Sigma)$ is the
space of Riemannian metrics on $\Sigma$.

A tensorial variable $A$ that is a functional of $h_{ij}$ and
$\pi^{ij}$ corresponds to a family of functions on the state space
$\Gamma$. The map $P: F(\Gamma) \rightarrow F(\Gamma)$, defined as
$P[A] = \langle A \rangle$ is a projector on $F(\Gamma)$ and can be
used to define the relevant variables with respect to the action of
the group $G$ on $\Gamma$.

We also note that  the map $\Pi: \Gamma \rightarrow \Gamma$, defined
as
\begin{eqnarray}
 \Pi[(h_{ij}, \pi^{ij})] = [\langle h_{ij} \rangle, \langle
\pi^{ij} \rangle], \label{proj}
\end{eqnarray}
projects onto the submanifold $\Gamma_0$ of $\Gamma$, which consists
of initial data invariant under the action $f$ of the group $G$. If
$G$ is the 6-dimensional group characterizing homogeneous and
isotropic spacetimes, $\Gamma_0$ consists of all constant curvature
metrics on $\Sigma$ and the corresponding conjugate momenta.

\subsection{Comments}
1. The construction above applies for a generic (compact) Lie group
$G$ and it does not specifically require isotropy and homogeneity.
For example, $G$ can be a group corresponding solely to special
homogeneity (Bianchi models). The only requirement is that the group
$G$ has a smooth action on the three-manifold $\Sigma$.
\\ \\
2. The  group $G$ was assumed compact. (When $G$ implements the
symmetry of homogeneity and isotropy,  $\Sigma$ also must be
compact: $\Sigma = S^3$.) However, the idea can also be applied to
non-compact groups, provided they possess a left-invariant measure
$d \mu[g]$: Let $O_n$ be a sequence of open subspaces of $G$ with
compact support, such that $O_{n-1} \subset O_n$ and $ \cup_n O_n =
G$. Then, for any tensor field $A^{i_1 \ldots i_n}{}_{j_1 \ldots
j_m}$ on $\Sigma$ we define
\begin{eqnarray}
\langle A^{i_1 \ldots i_n}{}_{j_1 \ldots j_m} \rangle (x) = \lim_{n
\rightarrow \infty} \frac{1}{\mu(O_n)} \int_{O_n} d \mu[g]
(f^*_gA)^{i_1 \ldots i_n}{}_{j_1 \ldots j_m}(x). \label{noncompact}
\end{eqnarray}

If this limit exists, and if it is independent of the choice of the
sequence $O_n$, then the group averaged tensor field $\langle A^{i_1
\ldots i_n}{}_{j_1 \ldots j_m} \rangle$ is invariant under the
action of the group $G$. The properties presented in Sec. 2.1 then
follow.

In the following, we shall assume that the group $G$ is compact,
keeping in mind that with suitable conditions the results can also
be applied to the non-compact case.
\\ \\
3. From the physical point of view there is an ambiguity in the
construction of the projector above. The action of a  group $G$ on
$\Sigma$ is unique {\em at most}  up to diffeomorphisms. If $f_g$ is
an action of $G$ on $\Sigma$, and $F$ a diffeomorphism (that does
not coincide with any of the diffeomorphisms $f_g$), then $F \circ
f_g \circ F^{-1}$ is a different group action on $\Sigma$.

Group-averaging is equivariant with respect to the action of the
diffeomorphism group $Diff(\Sigma)$, i.e.
\begin{eqnarray}
F^*\langle A^{i_1 \ldots i_n}{}_{j_1 \ldots j_m}\rangle = \langle
(F^*A)^{i_1 \ldots i_n}{}_{j_1 \ldots j_m} \rangle',
\end{eqnarray}
where $\langle \cdot \rangle'$ denotes the average with respect to
the group action $F \circ f_g \circ F^{-1}$.

Hence, if $\Gamma_0$ is the submanifold of initial data invariant
under the $f$-action of $G$ on $\Sigma$, $F^*\Gamma_0$ is the
submanifold invariant under the $F\circ f\circ F^{-1}$ action of
$G$. The effective description of the group-averaged quantities
 admits the group $Diff(\Sigma)$ as a gauge symmetry.
It is therefore necessary to reduce the system by removing the gauge
degrees of freedom corresponding to the $Diff(\Sigma)$-symmetry.
This is equivalent to the selection of a specific group action among
the class of diffeomorphic equivalent actions.  We shall see in the
next section, that the presence of a perfect fluid allows for a {\em
gauge-invariant} reduction of the $Diff(M)$ symmetry.

\subsection{The group averaging calculus}

 Let $A$ be a tensor field invariant under the action of
the group of isometries, i.e., $f^*_gA = A$ for all $g \in G$. Then
for any tensor field $B$
\begin{eqnarray}
\langle A \otimes B \rangle = \int d \mu[g] f_g^*(A\otimes B) = \int
d \mu[g] f_g^*A \otimes f_g^*B = \int d \mu[g] A \otimes f_g^*B
\nonumber \\ = A \otimes \int d \mu[g] f_g^*B = A \otimes  \langle B
\rangle. \label{idty}
\end{eqnarray}
This means that an invariant tensor can be taken out of the group
averaging operation. The  same holds if some indices of $A$ and $B$
are contracted.

\medskip

Another  important property of group averaging is that in a
spacetime characterized by homogeneity and isotropy, the group-
averaging of a scalar field  equals  the spatial average of the
field on the three-surface $\Sigma$. The proof is the following.

We consider  the case of compact three-surface $\Sigma$ and group
$G$. Let $G_x$ be the stability group of $x \in \Sigma$, i.e. the
subset of $G$ of all elements $g \in G$, such that $f_g(x) = x$. The
quotient $G/G_x$ coincides with $\Sigma$, so that $G$ forms a fiber
bundle over $\Sigma$ with fiber $G_x$. Considering a local
trivialization of the bundle $\phi_x: G \rightarrow \Sigma \times
G_x$, such that $\phi_x(g) = (f_g(x), g')$, where $g' \in G_x$. Then
the measure $d \mu(g)$ splits as $ d \mu_{\Sigma}(y) d
\mu_{G_x}(g')$, $y = f_g(x)\in \Sigma$, where $d \mu_{\Sigma}$ is
the G- invariant measure on $\Sigma$. Hence
\begin{eqnarray}
\langle \phi \rangle(x) = \int d \mu(g) \phi[f^{-1}_g(x)] = \int d
\mu_{\Sigma}(y) \phi(y) \left(\int d \mu_{G_x} \right) = c \int d
\mu_{\Sigma}(y) \phi(y), \label{scal}
\end{eqnarray}
where $c = \left(\int d \mu_{G_x} \right)$ is a constant (it does
not depend on $\phi$). Since for the constant function $\phi(x) =
1$, $\langle \phi \rangle(x) = 1$, $c = 1/V$, where $V$ is the
volume of $\Sigma$ with respect to the invariant metric. We have
therefore shown that the group average of a scalar function equals
its spatial average over $\Sigma$ with respect to the
group-invariant measure.

The {\em only} properties of group averaging we will use in this
paper are the identity (\ref{idty}) and group-averaging of scalar
fields. The reason is that we will only encounter group averages of
tensors with two indices, which, when contracted with the
group-invariant FRW metric $\bar{h}_{ij}$,  lead to group averages
of scalars. It turns out that these are the only averages that
appear explicitly in the backreaction equations.

\section{The Hamiltonian description of fluids}

The Hamiltonian formalism is well suited for dealing with the
problem of backreaction, because the level of description associated
with homogeneity and isotropy corresponds to a submanifold of the
canonical state space of a gravity theory. Moreover, in spacetimes
with a perfect fluid, the gauge symmetry of spatial diffeomorphisms
can be factored out completely, due to the special properties of the
perfect fluid's Lagrangian. Perfect fluid spacetimes suffice for
many cosmological applications. Dust-filled spacetimes, in
particular, are  relevant for examining the issue of
backreaction-induced cosmic acceleration.

In this section, we present the Lagrangian and Hamiltonian formalism
for relativistic gravitating perfect fluids. There exist several
different approaches \cite{pf}; here, we follow an adaptation of the
formalism presented in Ref. \cite{KSG}.

\subsection{Perfect fluid Lagrangian}

 \paragraph{Thermodynamics.} The thermodynamic properties of a fluid are encoded in
  the internal energy (Gibbs) functional $e (V,S)$, which expresses
the internal energy per particle $e$ as a function of the specific
volume $V$  and the specific entropy  $S$. The first law of
thermodynamics takes the form
\begin{eqnarray}
de = - P dV + T dS,
\end{eqnarray}
where $P$ is the pressure and $T$ the temperature.

We consider the special case that the internal energy is a function
of the specific volume $V$ only--i.e., we ignore thermal effects.
The internal energy can be written as $e(1/n)$, where $n = 1/V$ is
the number density. The energy density is $\rho = e/V = n e$. We
note that for an equation of state $P = w \rho$, the Gibbs
functional is
\begin{eqnarray}
e = c n^w,
\end{eqnarray}
where $c$ a constant. In particular, for $w = 0$ (dust), we obtain
$e = c  $, where $c$ has  dimension of mass and hence $\rho = c n$.

\paragraph{The matter space.} Let $Z$ be the {\em matter space},
i.e., a three-dimensional manifold, whose points correspond to
material particles. These particles are distinguishable, in the
sense that each point of $Z$ corresponds to a particle of definite
identity. The configuration of the fluid is fully determined, if,
for every point $X$ of the spacetime $M$, one specifies a particle
$z \in Z(X)$, whose worldline passes through $X$. Hence, a
configuration of the fluid is represented by an on-to mapping
$\zeta: M \rightarrow Z$. Given a coordinate system  in $Z$, the
mapping is described by three functions $\zeta^i(X)$.

Globally, if $M = \Sigma \times R$, where $\Sigma$ is a
three-manifold, $Z$ must be diffeomorphic to $\Sigma$. $Z$ should be
a homogeneous space, namely it should carry the transitive action of
a Lie group.  $Z$ is also  equipped with a volume three-form $\nu$
\begin{eqnarray}
 \nu = \nu(z)\; dx^1 \wedge dx^2 \wedge dx^3,
\end{eqnarray}
which measures the number of particles $n(D)$ within any given
region $D$ in $Z$: $n(D) = \int_D \nu$.

The pullback $\zeta^*\nu$ is a three-form on the spacetime $M$. It
is closed ($d (\zeta^*\nu) = 0$), since $d\nu = 0$ on the
three-dimensional manifold $Z$. This implies that the corresponding
vector density
\begin{eqnarray}
j^{\mu} = - \nu  \; \epsilon^{\mu \nu \rho \sigma}
\partial_{\nu}\zeta^1 \partial_{\rho}\zeta^2
\partial_{\sigma}\zeta^3, \label{jm0}
\end{eqnarray}
satisfies the conservation equation
\begin{eqnarray}
\partial_{\mu} j^{\mu} = 0. \label{conserv1}
\end{eqnarray}

The vector density $j^{\mu}$ is the conserved particle-number
current. It can be split into a normalized velocity vector field
$u_{\mu}$ and a particle number density $n$ as
\begin{eqnarray}
j^{\mu} = \sqrt{-g} n u^{\mu}, \label{um}
\end{eqnarray}

whence
\begin{eqnarray}
n = \frac{1}{\sqrt{-g} } \sqrt{ - j^{\mu} j^{\nu} g_{\mu \nu}}.
\label{numberdensity}
\end{eqnarray}

\paragraph{First-order Lagrangian.} Having expressed the particle
density as a function of the first derivatives of the fields
$z^i(X)$, we  write a first-order Lagrangian that reproduces the
fluid's equations of motion
\begin{eqnarray}
L = \sqrt{-g} \rho[n] = \sqrt{-g} n e[1/n].
\end{eqnarray}
The three variational equations of $L$ with respect to $\zeta^i$
together with the equation (\ref{conserv1}) lead to the conservation
equation for the stress-energy tensor \cite{KSG}
\begin{eqnarray}
\nabla_{\mu} T^{\mu \nu} = 0,
\end{eqnarray}
where
\begin{eqnarray}
T^{\mu \nu} = - \sqrt{-g}[\rho u^{\mu} u^{\nu} + P(g^{\mu \nu}  +
u^{\mu} u^{\nu})].
\end{eqnarray}
When the gravitational field is included, the total action
\begin{eqnarray}
S[g, \zeta] = \int d^4X \sqrt{-g} \left(\frac{1}{\kappa} R - \rho[n]
\right) \label{action}
\end{eqnarray}
leads to the Einstein equations $G^{\mu \nu} = \frac{\kappa}{2}
T^{\mu \nu}$, where $\kappa = 16 \pi G$.

\subsection{3+1 splitting}

We next perform a change of variables in the action (\ref{action}).
This change of variables corresponds to  a 3+1 splitting of
spacetime with respect to the matter space $Z$. Let $t(X)$ be a time
function on the spacetime $(M, g)$, i.e., a scalar function on $M$,
such that the condition $t(X) = t$ defines Cauchy surfaces
$\Sigma_t$, and $\cup_t \Sigma_t = M$.

We define a diffeomorphism from $M$ to $Z \times R$ as $X
\rightarrow [\zeta(X), t(X)]$. The inverse map ${\cal E}: Z \times R
\rightarrow M$, $(x^i, t) \rightarrow X^{\mu} = {\cal E}^{\mu}(X)$
is a functional of $\zeta$ and allows for the standard 3+1
decomposition of the spacetime metric $g_{\mu \nu}$.

Using the vector fields
\begin{eqnarray}
t^{\mu}(X)  &=& \frac{\partial {\cal E}^{\mu}}{\partial t} [t(X),
\zeta(X)] \\
{\cal E}^{\mu}_i (X)&=& \frac{\partial {\cal E}^{\mu}}{\partial x^i}
[t(X), \zeta(X)]
\end{eqnarray}

we  express the Riemannian three-metric $h_{ij}$ on $\Sigma_t$
(obtained from the pull-back of $g_{\mu \nu}$ to $\Sigma_t$ under
${\cal E}$) as
\begin{eqnarray}
h_{ij}(t,x) = {\cal E}^{\mu}_i(t,x) {\cal E}^{\nu}_j(t,x) g_{\mu
\nu}[{\cal E}(t,x)].
\end{eqnarray}

Let $n^{\mu}$ be the unit normal on the surfaces $\Sigma_t$; then
$n_{\mu} {\cal E}^{\mu}_i = 0$. The lapse function $N$ and the shift
vector $N^i$ are defined as
\begin{eqnarray}
t^{\mu} = N n^{\mu} + {\cal E}^{\mu}_i N^i. \label{decomp}
\end{eqnarray}
The extrinsic curvature tensor on $\Sigma_t$ equals
\begin{eqnarray}
K_{ij} := {\cal E}^{\mu}_i {\cal E}^{\nu}_j \nabla_{\mu} n_{\nu} =
\frac{1}{2N} (\dot{h}_{ij} - {}^3\nabla_i N_j - {}^3\nabla_j N_i),
\end{eqnarray}
with ${}^3 \nabla_i$ the covariant derivative on $Z$ compatible with
the three-metric $h_{ij}$--the dot denotes derivative with respect
to $t$. We also note that $\sqrt{-g} = N \sqrt{h}$, where $h = det
h_{ij}$.

The current density $j^{\mu}$ of Eq. (\ref{jm0}) takes the form
\begin{eqnarray}
j^{\mu} = \nu t^{\mu}. \label{jm}
\end{eqnarray}
Substituting Eq. (\ref{jm}) into Eq. (\ref{um}), we obtain
\begin{eqnarray}
n = \nu \frac{\sqrt{ 1 - h_{ij}\tilde{N}^i \tilde{N}^j}}{\sqrt{h}},
\end{eqnarray}

where $\tilde{N}^i = N^i/N$.

 The action
(\ref{action}) then equals (up to boundary terms)
\begin{eqnarray}
S = \int dt \int d^3z N \frac{\sqrt{h}}{\kappa} \left(K_{ij} K^{ij}
- K^2 + {}^3R - \kappa n e[n]\right), \label{action2}
\end{eqnarray}
where $K = K_{ij}h^{ij}$ and ${}^3R$ is the Ricci scalar on $Z$
associated to  the three-metric $h_{ij}$.

The action (\ref{action}) is a functional on a space $Q$ spanned by
the 13 independent field variables $g_{\mu \nu}, \zeta^i$. However,
after the redefinition of the variables involved in the 3+1
splitting, the action takes the form (\ref{action2}) that depends on
the 10 variables $h_{ij}, N^i, N$. The change of variables allowed
for the separation  of the physical from the gauge degrees of
freedom. The action (\ref{action2}) is defined on the space
$Q_{red}$ spanned by the field variables $(h_{ij}, N^i, N)$,

This change of variables  is physically equivalent to the imposition
of the comoving coordinate conditions. However, {\em no gauge-fixing
was involved}. The reduction took place at the Lagrangian level: the
variables $h_{ij}, N, N^i$ are {\em functionals} of the original
variables $g_{\mu \nu}, \zeta^i$. They are  coordinates on the space
$Q$, invariant, by construction, under the `gauge symmetry' of
spatial diffeomorphisms. It follows that (at least locally)
$Q_{red}$ can be identified with the quotient space $Q/
Diff(\Sigma)$.

The construction above does not apply to generic matter Lagrangians:
the perfect fluid Lagrangian is special in that it allows a
gauge-invariant way to select a spatial coordinate system tied on
the matter degrees of freedom.

\subsection{The Hamiltonian description}

We next perform the Legendre transform of the action
(\ref{action2}). The momenta conjugate to $h_{ij}$ are
\begin{eqnarray}
\pi^{ij} = \frac{\sqrt{h}}{\kappa}(K^{ij} - K h^{ij}),
\end{eqnarray}
while the variables conjugate to $N$ and $N_i$ vanish identically
(they correspond to primary constraints).

The Legendre transform yields the Hamiltonian
\begin{eqnarray}
H = \int d^3z  N \left({\cal H} + \tilde{N}^i {\cal H}_i +
\sqrt{h}\, \rho\left[\nu \frac{\sqrt{ 1 - h_{ij}\tilde{N}^i
\tilde{N}^j}}{\sqrt{h}}\right] \right), \label{Ham0}
\end{eqnarray}
where
\begin{eqnarray}
{\cal H} &=& \frac{\kappa }{\sqrt{h}} (\pi^{ij}\pi_{ij} -
\frac{1}{2}
\pi^2) - \frac{\sqrt{h}}{\kappa} \,{}^3 R \label{superhamiltonian}\\
{\cal H}^i &=& - 2 \nabla_j \pi^{ij}. \label{supermomentum}
\end{eqnarray}
From now on we drop the prefix ${}^3$ from the spatial covariant
derivative.

Variation with respect to $\tilde{N}^i$ yields
\begin{eqnarray}
{\cal H}^i = \frac{\tilde{N^i}}{\sqrt{1 - \tilde{N}^i \tilde{N}_i}}
\; \nu \rho',
\end{eqnarray}
which can be solved for $\tilde{N}^i$:
\begin{eqnarray}
\tilde{N}^i = \frac{{\cal H}^i}{\sqrt{\nu^2 F^2 + {\cal H}^i{\cal
H}_i}}, \label{tilden}
\end{eqnarray}

where $F(x)$ is a function of  $x= {\cal H}^i {\cal H}_i/\nu^2$
defined implicitly by the algebraic equation
\begin{eqnarray}
F(x) = \rho' \left( \frac{\nu}{\sqrt{h}} \frac{1}{\sqrt{1 +
x/F(x)^2}} \right).
\end{eqnarray}

Substituting Eq. (\ref{tilden}) back to the Hamiltonian we obtain
\begin{eqnarray}
H = \int d^3z N \left[{\cal H} + \frac{{\cal H}_i {\cal
H}^i}{\sqrt{\nu^2 F^2 + {\cal H}^i {\cal H}_i}} + \sqrt{h} \, \rho
\left( \frac{\nu/\sqrt{h}}{\sqrt{1 + {\cal H}^i {\cal
H}_i/\nu^2F^2}}\right) \right] \label{Ham1}
\end{eqnarray}

Variation with respect to the lapse $N$ yields the constraint
equation
\begin{eqnarray}
{\cal H} + \frac{{\cal H}_i {\cal H}^i}{\sqrt{\nu^2 F^2 + {\cal H}^i
{\cal H}_i}} + \sqrt{h} \rho \left( \frac{\nu/\sqrt{h}}{\sqrt{1 +
{\cal H}^i {\cal H}_i/\nu^2F^2}}\right) = 0 . \label{constr}
\end{eqnarray}

The gravitating perfect fluid with the variables above is then a
first-class constrained system, with the Hamiltonian vanishing on
the constraint surface. We see that the use of coordinates tied on
the matter state have led effectively to an automatic implementation
of the spatial diffeomorphism constraints. Active diffeomorphisms
are not gauge symmetries of the system described by Eqns.
(\ref{Ham1} and (\ref{constr}).

For a dust-filled spacetime ($\rho = c n$), we obtain $F(x) = c$ and
the Hamiltonian simplifies:
\begin{eqnarray}
H = \int d^3z N \left[{\cal H} + \sqrt{\mu^2 + {\cal H}^i {\cal
H}_i}\right], \label{Ham2}
\end{eqnarray}
where $\mu = c \nu$.

\section{The effective description}

\subsection{The symmetry surface}

Having obtained the Hamiltonian for a gravitating perfect fluid, we
 proceed to the description of the effective dynamics in
terms of a homogeneous and isotropic metric.

The Hamiltonian (\ref{Ham1}) is defined on the space $\Gamma =
T^*Riem(Z)$ spanned by the variables $h_{ij}(x)$ and $\pi^{ij}(x)$.
$\Gamma$ carries a symplectic form
\begin{eqnarray}
\Omega = \int_Z d^3 x \,  \delta \pi^{ij}(x) \wedge \delta
h_{ij}(x).
\end{eqnarray}

For concreteness, we set $Z = S^3$. Let ${}^0h_{ij}$ be the metric
of the unit sphere
\begin{eqnarray}
{}^0h_{ij}dx^i dx^j= d \chi^2 + \sin^2 \chi (d \theta^2 + \sin^2
\theta d \phi^2).
\end{eqnarray}

This metric is our standard for isotropy and homogeneity in the
matter space. It has six linearly independent Killing vectors. These
vectors generate an action $f$ of the group $G$ on $Z$. This action
extends to an action on the space $Riem(Z)$ of three-metrics on $Z$
and on the phase space $\Gamma$. We will employ this group action in
order to define group averaging for tensors on $Z$ as described in
section 2.

Let $\Gamma_0$ be the submanifold of $\Gamma$ invariant under the
action $f$; we will refer to $\Gamma_0$ as the {\em symmetry
surface}. It consists of pairs $(\bar{h}_{ij}, \bar{\pi}^{ij})$ of
the form $\bar{h}_{ij} = \alpha^2 \; {}^0h_{ij}$ and $\bar{\pi}^{ij}
= \frac{p}{12 \pi^2 \alpha} \sqrt{{}^0h}\; {}^0h_{ij}$. $\Gamma_0$
is a symplectic submanifold of $\Gamma$, in which the variables
$\alpha$ and  $p$ define a symplectic chart, i.e., the restriction
of $\Omega$ on $\Gamma_0$ is
\begin{eqnarray}
\Omega_{\Gamma_0} = dp\wedge d\alpha.
\end{eqnarray}

We also assume that the number density three-form $\nu$ is invariant
under the group action on $Z$. This implies that $\nu = \nu_0
\sqrt{{}^0h}$, where $\nu_0$ is a constant.

The submanifold $\Gamma_0$  defines the level of description in
terms of homogeneous and isotropic variables. We shall project the
dynamics of the gravitating fluid onto  $\Gamma_0$. The group action
$f$ allows us to define group averaging for the variables $h_{ij},
\pi^{ij}$ through Eq. (\ref{aver}), and thus to construct a
projector $\Pi:\Gamma \rightarrow \Gamma_0$, such that
\begin{eqnarray}
\Pi (h_{ij}, \pi^{ij}) = (\langle h_{ij} \rangle, \langle \pi^{ij}
\rangle).
\end{eqnarray}

Hence, the coordinates of any   point of $\Gamma$ split uniquely as
\begin{eqnarray}
(h_{ij}, \pi^{ij})  = (\bar{h}_{ij} + \delta h_{ij}, \bar{\pi}^{ij}
+\delta \pi^{ij}),
\end{eqnarray}
where $\bar{h}_{ij} = \langle h_{ij}\rangle, \bar{\pi}^{ij} =
\langle \pi^{ij} \rangle$ and $\langle \delta h_{ij} \rangle = 0,
\langle \delta \pi^{ij} \rangle = 0 $.

\subsection{Choice of time variable}

In order to write the evolution equations corresponding to the
Hamiltonian (\ref{Ham1}) and the constraint (\ref{constr}), one
needs to choose a time variable, i.e., to specify the time function
$t(X)$ employed in the 3+1 decomposition of the Lagrangian---see
Sec. 3.2. To do so we must identify a family of timelike curves
spanning the spacetime $M$, such that $t$ coincides with their
proper time. The equations of motion obtained from the Hamiltonian
(\ref{Ham1}) hold for any choice of time, as long as we take into
account that the definition of the canonical data depends explicitly
in the choice of the time function\footnote{This means that the time
function $t(X)$ must be a functional of the configuration space
variables $g_{\mu \nu}$ and $\zeta^i$. Hence, the foliation defined
from $t(X)$ and $\zeta(X)$ is a functional of the configuration
variables. Such "foliation functionals" have been introduced in Ref.
 \cite{Sav}. The specific functional corresponding to time measured by comoving observers
 is equivariant with respect to spacetime diffeomorhisms. }.

The choice of time variable must be made consistently over the whole
gravitational phase space. This is made possible by the presence of
a perfect fluid, because there is a natural choice of time
associated to comoving observers. Comoving observers correspond to
the integral curves of the fluid's four-velocity $u^{\mu}$---see Eq.
(\ref{um}). In matter coordinates, Eq. (\ref{um}) implies that
\begin{eqnarray}
u^{\mu} = \frac{t^{\mu}}{\sqrt{-t^{\mu} t_m}} = \frac{1}{N \sqrt{1 -
\tilde{N}_i \tilde{N_i}}} \left(\frac{\partial}{\partial
t}\right)^{\mu}.
\end{eqnarray}
The parameter $t$  corresponds to the proper time along the comoving
observers only if
\begin{eqnarray}
N = \frac{1}{\sqrt{1 - \tilde{N}^i\tilde{N}_i}}.
\end{eqnarray}

The decomposition (\ref{decomp}) of $t^{\mu}$ implies that $N^i =
u^i$, where $u^i$ is the spatial component of the three-velocity for
the comoving observers; $u^i$ corresponds to the fluid-particles'
deviation velocity from the uniform cosmological expansion.
$\tilde{N}^i$ coincides with $v_i = u^i/ \sqrt{1 + u^k u_k}$,
namely, the three-velocity of observers along worldlines normal to
the hypersurfaces $\Sigma_t$.

 Eq. (\ref{tilden}) implies that
\begin{eqnarray}
u^i = \frac{{\cal H}^i}{\nu F({\cal H}^k {\cal H}_k/\nu^2)}.
\end{eqnarray}

We note  that for a FRW spacetime ${\cal H}^i = 0$; hence, $u_i =
0$: on a homogeneous and isotropic spacetime, the worldlines of the
comoving observers are normal to the homogeneous and isotropic
hypersurfaces. Moreover, they are geodesics (since $N=1$).

We next consider a spacetime that is not homogeneous and isotropic.
The non-vanishing of the vector field $u_i$ measures the degree of
deviation of the comoving observers   from being normal to the
homogeneous and isotropic hypersurfaces and   from being geodesics.

Test particles move on geodesics of the spacetime metric. This
follows under the assumption that the particle's backreaction to the
geometry is insignificant. However, when  we consider a large number
of particles, we have to take their mutual interaction into account.
This results into `forces': the motion of an individual particle is
no more a free-fall, and as such it does not correspond to a
geodesic of the metric.

When we model an inhomogeneous spacetime by an FRW solution, we
essentially assume that the world-lines of the particles are
geodesics. Hence, we ignore the `inter-particle interaction' and we
assume that individual particles evolve under the average fields
generated by the whole matter content of the universe. In this
sense, the FRW description is conceptually  similar  to the mean
field approximation in statistical mechanics. To move beyond the
mean field approximation, it is necessary to consider fluctuations.
In many-body systems, the effect of fluctuations
 is substantial if they exhibit strong (long-range) correlation. In analogy,
 the incorporation of back-reaction in the
 cosmological setting might result in a substantial
 change over the FRW predictions if there are persistent
 correlations between the properties of the inhomogeneous regions.

\subsection{Equations of motion}

Having chosen the time variable $t$ to be the one measured by
comoving observers, we write now the equations of motion following
by the Hamiltonian (\ref{Ham0}). We recall  that with this choice
$N^i = u^i = {\cal H}^i/\nu F({\cal H}^k {\cal H}_k/\nu^2)$ and $N =
\sqrt{1 + u^iu_i}$.

\begin{eqnarray}
\dot{h}_{ij} &=& \frac{2N \kappa}{\sqrt{h}} (\pi_{ij} - \frac{1}{2}
\pi h_{ij}) + {\cal L}_u h_{ij}  \label{HE1} \\
\dot{\pi}^{ij} &=& - \frac{N  \sqrt{h}}{\kappa} (R^{ij} -
\frac{1}{2} R h^{ij}) - \frac{2 N \kappa}{\sqrt{h}} (\pi^{ik}
\pi_k{}^j - \frac{1}{2} \pi \pi^{ij}) - \frac{N \kappa}{2 \sqrt{h}}
h^{ij}(\pi^{kl} \pi_{kl} + \frac{1}{2} \pi^2) \nonumber \\
&+& \sqrt{h} (\nabla^i \nabla^jN - h^{ij} \nabla^k \nabla_k N) +
{\cal L}_u \pi^{ij} \nonumber \\&+& \frac{1}{2} N \sqrt{h} \{h^{ij}
P(n) + u^i u^j [\rho(n) - P(n)]\} , \label{HE2}
\end{eqnarray}

where $P = \rho - n \frac{\partial \rho}{\partial n}$ is the
pressure, $n = \frac{\nu}{\sqrt{h}}/\sqrt{1 + u^k u_k}$ and the Lie
derivatives ${\cal L}_u$ read explicitly
\begin{eqnarray}
{\cal L}_u h_{ij} &=& \nabla_i u_j +\nabla_j u_i \\
{\cal L}_u \pi^{ij}  &=& \pi^{i k} \nabla_k u^j + \pi^{kj} \nabla_k
u^i - u^k \nabla_k \pi^{ij} - \pi^{ij} \nabla_k u^k,
\end{eqnarray}

The initial data satisfy the constraint equation
\begin{eqnarray}
\Phi = \frac{\kappa}{h} (\pi^{ij} \pi_{ij} - \frac{1}{2} \pi^2) -
\frac{1}{\kappa} R  + [ u^k u_k (\rho(n) - P(n)) + \rho(n)] = 0 .
\label{constraintf}
\end{eqnarray}

Note that $\Phi$ is a scalar on $\Sigma$.

\paragraph{Symmetric solutions} The symmetry surface $\Gamma_0$ is invariant under the
Hamiltonian evolution. Setting $h_{ij} = \alpha^2 \; {}^0 h_{ij}$
and $\pi^{ij} = \frac{p}{12\pi^2 \alpha} \sqrt{{}^0h} \;
{}^0h^{ij}$, we obtain $u^i = 0 , N = 1$ and reduced Hamilton
equations,
\begin{eqnarray}
\dot{\alpha} &=& - \kappa \frac{p}{24 \pi^2 \alpha} \\
\frac{d}{dt}(\frac{p}{\alpha}) &=& (\kappa \frac{p^2}{48 \pi^2
\alpha^3} + \frac{2 \pi^2}{\kappa \alpha} + 6 \pi^2 P \alpha)
\end{eqnarray}
together with the constraint $\kappa\frac{p^2}{96 \pi^4 \alpha} +
\frac{6 \alpha}{\kappa}  - \rho \alpha^3 = 0$.
 The usual form of the FRW equations then follows
\begin{eqnarray}
\left(\frac{\dot{\alpha}}{\alpha}\right)^2 + \alpha^{-2} = \frac{\kappa}{6} \rho \\
\frac{\ddot{\alpha}}{\alpha} = - \frac{\kappa}{12} (\rho + 3 P).
\label{FRW2}
\end{eqnarray}

\subsection{Effective equations}

We next construct the effective evolution equations for homogeneous
and isotropic metrics $\bar{h}_{ij}$ and conjugate momenta
$\bar{\pi}^{ij}$. This is equivalent to a projection of the Hamilton
equations from the full phase space to the symmetry surface
$\Gamma_0$.   Let the system be at point $(h_{ij}, \pi^{ij})$ of the
constraint surface at time $t$. The corresponding relevant variables
are $(\bar{h}_{ij}, \bar{\pi}^{ij}) = (\langle h_{ij} \rangle,
\langle \pi^{ij}\rangle)$; they define a point of the symmetry
surface. At time $t + \delta t$, the system lies at $(h_{ij} +
\delta t \dot{h}_{ij}, \pi^{ij} + \delta t \dot{\pi}^{ij})$. The
effective equation of motion is obtained by projecting $(h_{ij} +
\delta t \dot{h}_{ij}, \pi^{ij} + \delta t \dot{\pi}^{ij})$ to
$\Gamma_0$. Hence,
\begin{eqnarray}
\frac{d}{d t} \bar{h}_{ij} = \langle \dot{h}_{ij} \label{be1}
\rangle \\
\frac{d}{d t} \bar{\pi}^{ij} = \langle \dot{\pi}^{ij} \rangle,
\label{be2}
\end{eqnarray}
where $\dot{h}_{ij}$ and $\dot{\pi}^{ij}$ are given by (\ref{HE1})
and (\ref{HE2}) respectively.

Equations (\ref{be1}-\ref{be2}) are exact: we  made no assumptions
about the form of the canonical variables $h_{ij}$, $\pi^{ij}$. In
order to derive  a meaningful effective description, we must specify
a region in the system's phase space. This is equivalent to a choice
of an approximation scheme, according to which the right-hand-side
of Eqs. (\ref{be1}-\ref{be2}) will be evaluated.

Here, we employ the following approximations.
\\ \\
{\bf 1. Perturbation expansion.} We first effect of a perturbation
expansion around the symmetry surface. We point out that {\em we do
not perturb around the solutions of the FRW equations}, for in this
case it would be impossible to obtain consistent backreaction
equations.

The first-order terms in the perturbation expansion vanish. This is
easy to see for ultra-local terms in the perturbations (for example
ones that involve $\langle \delta h_{ij} \rangle$), because these
vanish by definition. There are however non-ultralocal first-order
terms, such as $\bar{\nabla}_k \delta h_{ij}$, which do not vanish
identically. It turns out that the contribution of all such terms
reduces to integrals over a total divergence and hence vanishes for
compact spacetimes.

(This property follows from the high degree of symmetry of the FRW
cosmologies. If, for example, we had considered expansion around the
symmetry surface for a three-dimensional isometry group, the
first-order contributions would be non-vanishing.)

Hence, the dominant terms in the back-reaction are second order to
the perturbations. We note that to second order $N \simeq 1 +
\frac{1}{2} \bar{h}^{ij}u_i u_j$, i.e.,  deviation velocities are
"non-relativistic".
\\ \\
{\bf 2. Ignore ultralocal terms.} . The second assumption in our
derivation of the backreaction equations is  that the fluctuations
$\delta h_{ij}(z)$ and $\delta \pi^{ij}(z)$ are small with respect
to the averages $\bar{h}_{ij}(z)$ and $\bar{\pi}^{ij}(z)$ in an {\em
ultralocal} sense. By this we mean that the trace  norm of, say, the
matrix $\delta h_{ij}(z)$  must be a fraction $\epsilon << 1$ of the
trace norm of the matrix $\bar{h}_{ij}(z)$ at almost all points of
$z$. Note that there is no ambiguity in the notion of the norm,
because $h_{ij}(x)$ is defined with reference to the comoving
co-ordinate system and  there is no "gauge" freedom from active
diffeomorphisms. However, the perturbations are functions of $x$: if
their characteristic scale is of order $l << 1$,\footnote{The
coordinates on $Z$ are dimensionless corresponding to the unit
sphere. In terms of distances, the condition $l << 1$ translates to
 fluctuations at a scale much smaller than the scale factor
$\alpha$.} the action of a derivative operator leads to a term of
order $\epsilon/l >> \epsilon$. This implies that the ultralocal
terms in the backreaction equation are much smaller than the contain
derivatives of $\delta h_{ij}, \delta \pi^{ij}$. We therefore drop
all terms that involve {\em only} ultralocal terms of the
perturbations.
\\ \\
Using the approximations above, we  consider  the special case of a
dust-filled spacetime $ \rho = c n$, where $c$ a constant. We write
$\mu = c \nu$, and $\mu_0 = c \nu_0$, so that $\mu = \mu_0
\sqrt{{}^0h}$; hence $\rho = c (\mu /\sqrt{h}) \sqrt{1 + u_k u^k}$.
The velocity field $u_i$ satisfies $u_i = {\cal H}_i/\mu$.

We define the tensor field $C^{i}_{{}jk}$ through the equation
\begin{eqnarray}
\nabla_i V^j - \bar{\nabla}_i V^j = C^j_{{}ik} V^k,
\end{eqnarray}
where $\bar{\nabla}_i$ is the covariant derivative associated to
$h_{ij} = \langle h_{ij}\rangle$. From this definition, we obtain
\begin{eqnarray}
C^{i}_{{}jk} = \frac{1}{2} \bar{h}^{il} \left(\bar{\nabla}_j \delta
h_{il} + \bar{\nabla}_k \delta h_{jl} - \bar{\nabla}_l \delta h_{jk}
\right).
\end{eqnarray}

We shall also use the variables
\begin{eqnarray}
\lambda^i :&=&  \bar{h}^{kl} C^i{}_{kl} = - \frac{1}{\sqrt{\bar{h}}}
\bar{\nabla}_k (\sqrt{h} h^{ik}), \label{lambdai} \\
\kappa_i : &=& C^j{}_{ji}. \label{kappa}
\end{eqnarray}

With the definitions above, ${\cal H}^i = \bar{\nabla}_j\delta
\pi^{ij} + \pi^{kl}C^i_{{}kl}$. Since a term $\delta \pi^{ij}
C^i_{{}jk}$ is of a higher order and does not appear in second-order
expansion
\begin{eqnarray}
{\cal H}^i = -2(\bar{\nabla}_j\delta \pi^{ij} +
\bar{\pi}^{kl}C^i_{{}kl}) = -2(\bar{\nabla}_j\delta \pi^{ij} +
\frac{p}{12\pi^2\alpha^2} \sqrt{\bar{h}} \lambda^i). \label{calH}
\end{eqnarray}

Expanding Eqs. (\ref{be1}-\ref{be2}) to leading order (second) in
perturbations around the symmetry surface and dropping the
sub-dominant  ultralocal terms we obtain
\begin{eqnarray}
\frac{d}{dt} \alpha &=& -\frac{\kappa p}{24 \pi^2 \alpha} (1 +
\frac{1}{2} \overline{u^2}) - \frac{1}{3} \Gamma  \alpha \label{be1a} \\
\frac{d}{dt} \left(\frac{p}{\alpha}\right) &=& \frac{\kappa p^2}{48
\pi^2 \alpha^3} (1 + \frac{1}{2} \overline{u^2}) + \frac{12
\pi^2}{\kappa \alpha} ( 1 + \frac{1}{2} \overline{u^2}) \nonumber\\
&+& \frac{2 \pi^2 \alpha}{\kappa}  \Delta   - 2 \pi^2 \frac{\mu_0}{
\alpha^2} \overline{u^2} - \frac{2}{3} \Gamma \frac{p}{\alpha},
\label{be2a}
\end{eqnarray}

The quantity $\Delta$ is obtained from the  perturbations of the
scalar curvature and equals
\begin{eqnarray}
\Delta =  \bar{h}_{ij} \langle \sqrt{h} (R^{ij} - \frac{1}{2}
h^{ij}R) \rangle.
\end{eqnarray}

The quantity $\overline{u^2}$ is the mean-square deviation velocity
\begin{eqnarray}
\overline{u^2} = \bar{h}^{ij} \langle u_i u_j \rangle,
\end{eqnarray}
and
\begin{eqnarray}
\Gamma := \langle \lambda^i u_i \rangle,
\end{eqnarray}

plays the role of a time dependent "dissipation" coefficient.

We  also average over the constraint Eq. (\ref{constraintf}).
Keeping only the second-order non ultralocal terms, we obtain
\begin{eqnarray}
\frac{\kappa p^2}{96 \pi^4 \alpha^4} + \frac{6  }{\kappa \alpha^3} +
\frac{1}{\kappa} \overline{\delta R} - \frac{\mu_0}{\alpha^3} ( 1 +
\frac{1}{2} \overline{u^2}) = 0, \label{constrain0}
\end{eqnarray}

In Eq. (\ref{constrain0}) $\overline{\delta R}$ is the perturbation
of the Ricci scalar $\overline{\delta R} = \langle R \rangle -
\bar{R}$, where $\bar{R}$ is the Ricci scalar associated to the
metric $\bar{h}_{ij}$. We find,

\begin{eqnarray}
\overline{\delta R} = \langle \kappa^i \kappa_i - C^{ijk}C_{kij}
\rangle.
\end{eqnarray}

Eqs. (\ref{be1a}, \ref{be2a}, \ref{constrain0}) form a system of
equations for the effective evolution on the symmetry surface
$\Gamma_0$. These equations depend on four functions of time
$\Gamma(t), \overline{u^2}(t), \Delta(t), \overline{\delta R}(t)$,
whose form depends explicitly on the state of the non-relevant
degrees of freedom; they are also functionally dependent on the
relevant variables $\alpha(t)$ and $p(t)$. Since we have three
differential equations determining the motion in a two-dimensional
surface, one of them can be viewed as a compatibility condition
between the four functions of time. It is convenient to use Eq.
(\ref{be2a}) for the determination of the independent function
$\Delta(t)$. Then the effective dynamics on $\Gamma_0$, can be
described by Eqs. (\ref{be1a}) and (\ref{constrain0}). These take
the form

\begin{eqnarray}
\dot{\alpha} = \xi (1 + \frac{1}{2} \overline{u^2}) - \frac{1}{3}
\Gamma \alpha \label{be1b} \\
\xi^2 + 1 + \frac{\alpha^2 \overline{ \delta R}}{6} = \frac{\kappa
\mu_0}{6 \alpha} (1 + \frac{1}{2} \overline{u^2}).
\label{constrain4}
\end{eqnarray}

where we introduced
\begin{eqnarray}
\xi = -\frac{\kappa p}{24 \pi^2 \alpha}.
\end{eqnarray}

In order to study the properties of the backreaction Eqs.
(\ref{be1b}) and (\ref{constrain4}), we must find explicit forms for
the functions $\Gamma(t), \overline{u^2}(t)$ and $\overline{\delta
R}(t)$. We explore different regimes for them in the following
sections.

\section{The regime of small perturbations}

\subsection{Time evolution of the backreaction parameters}

Eqs. (\ref{be1b}) and (\ref{constrain4}) provide the evolution of
the relevant (FRW) variables including the backreaction terms. They
are valid at each moment of time. However, they are not, as yet,
dynamical equations. The reason is that the coefficients $\Gamma,
\overline{u^2}$ and  $\overline{\delta R}$ that appear in these
equations depend on the variables $\alpha$ and $\xi$. We must find
their explicit functional relation, in order to
 obtain a closed set of dynamical equations for the level of
description.

To compute the variables $\Gamma = \langle \lambda^i u_i \rangle$
and $\overline{u^2} = \langle u_ku^k \rangle$, it is necessary to
find the evolution equations for $\lambda^i $ and $u_i$. In this
section, we do so, to leading (second) order in the perturbations,
ignoring the ultralocal terms.

 \paragraph{Evolution of $\overline{u^2}$.} Using Eqs. (\ref{supermomentum}), (\ref{Ham2}) and the fact that
$u_i = {\cal H}_i/\mu$ we find that on the constraint surface
\begin{eqnarray}
\dot{u}_i = - \nabla_i (u_ku^k). \label{ui}
\end{eqnarray}
Eq. (\ref{ui}) is {\em exact}, i.e., no approximation has been
employed in its derivation. We note that the right-hand-side in Eq.
(\ref{ui}) is of higher order to the perturbations than the
left-hand-side. Hence, to leading order $u^i$ is a constant. We do
not expect this to be the case in general; however, this result
suggests that for sufficiently small deviations from the FRW
evolution, the deviation velocity vary slowly with time.

To leading-order in the perturbations $\overline{u^2}  =
\bar{h}^{ij}(t) \langle u_i u_j \rangle $, and since $u_i$ is
constant
\begin{eqnarray}
\overline{u^2}(t) = \left(\frac{\alpha(t_0)}{\alpha(t)}\right)^2
\overline{u^2}(t_0). \label{u2t}
\end{eqnarray}

\paragraph{Evolution of $\Gamma$.}
In order to calculate $\lambda^i$, we use Eqs. (\ref{lambdai}) and
(\ref{HE1}). We obtain
\begin{eqnarray}
\dot{\lambda}^i = - 3 \frac{\dot{\alpha}}{\alpha} + \frac{2 \kappa
}{\sqrt{\bar{h}}} \bar{\nabla}_k \pi^{ki} - \frac{\kappa}{2
\sqrt{\bar{h}}} \bar{\nabla}_k (\pi h^{ki}) +
\frac{1}{\sqrt{\bar{h}}} \bar{\nabla}_k[\sqrt{h}( \nabla^i u^k +
\nabla^k u^i - h^{ik} \nabla_l u^l)]. \label{ldot1}
\end{eqnarray}

Keeping  first-order perturbation terms around the symmetry surface
 and using Eq. (\ref{calH}), we obtain
\begin{eqnarray}
\dot{\lambda}^i = -3 \frac{\dot{\alpha}}{\alpha} \lambda^i - \kappa
\frac{\mu u^i}{\sqrt{\bar{h}}} - \frac{\kappa p}{24 \pi^2 \alpha^2}
\lambda^i - \frac{\kappa}{2} \bar{\nabla}^i(\pi/\sqrt{h}) +
\bar{\nabla}_k (\bar{\nabla}^i u^k + \bar{\nabla}^k u^i -
\bar{h}^{ki} \bar{\nabla}_l u^l ). \label{ldot2}
\end{eqnarray}
Then, to second order in perturbations
\begin{eqnarray}
\dot{\Gamma} =  \frac{\xi - 3 \dot{\alpha}}{\alpha} \Gamma - \kappa
\bar{\rho} \overline{u^2}+   \frac{\kappa}{2} \psi + 3
\overline{\Theta^2} \label{gammadot}
\end{eqnarray}
where  we wrote
\begin{eqnarray}
\psi = \langle \frac{\pi}{\sqrt{h}} \bar{\nabla}_ku^k \rangle, \\
\overline{\Theta^2} = \frac{1}{3}  \langle(\Theta^k_k)^2 - 2
\Theta_{ij} \Theta^{ij}\rangle, \label{pi2}
\end{eqnarray}
and
\begin{eqnarray}
\bar{\rho} = \mu_0/\alpha^3
\end{eqnarray}
is the matter density of the corresponding homogeneous and isotropic
solution.

The tensor $\Theta_{ij}$ is defined as
\begin{eqnarray}
\Theta_{ij} = \frac{1}{2} \left(  \bar{\nabla}_i u_j +
\bar{\nabla}_j u_i \right). \label{piij}
\end{eqnarray}

$\Theta_{ij}$ is  the expansion tensor for the (typically
non-geodesic) congruence of comoving observers, defined with respect
to the homogeneous and isotropic metric.

In linear response theory for non-relativistic fluids, the tensor
$\Theta_{ij}$ is proportional to the fluid's pressure tensor
\cite{DGM}. This suggests that $\Theta_{ij}$ incorporates the
effects of pressure generated by the "gravitational interaction" of
the inhomogeneities. It vanishes  for a homogeneous and isotropic
solution.

In the regime that $u_i$ is constant, $\Theta_{ij}$ is constant,
too. It follows that to  leading-order in perturbations
\begin{eqnarray}
\overline{\Theta^2} (t) = \left(\frac{\alpha(t_0)}{\alpha(t)}
\right)^4 \overline{\Theta^2}(t_0). \label{p2t}
\end{eqnarray}

 Eq. (\ref{gammadot}) is a differential equation for $\Gamma$
in terms of  $\overline{u^2}$ and $\overline{\Theta^2}$, which are
known functions of time, $\frac{\dot{\alpha}}{\alpha}$ and
$\xi/\alpha$, which is described by the evolution equations and
$\psi$, which is, as yet, undetermined. In effect, $\psi$  does not
allow for the existence of a closed system of equations at our level
of description. The derivation of a closed system of equations
requires additional assumptions.

To this end, we consider different types of perturbations and how
the evolution of $\Gamma$ differs according to the type.

\bigskip

{\em a. Dominant conformal perturbations.} In this regime, the
perturbations of the metric and conjugate momentum are of the form
$\delta h_{ij} \simeq b \bar{h}_{ij}$ and $\delta \pi^{ij} \simeq b'
\sqrt{\bar{h}} \bar{h}^{ij}$, for some scalar functions $b$ and
$b'$. Then, the term $\bar{\nabla}_k (\pi h^{ik})$ in Eq.
(\ref{ldot2}) equals $3 \bar{\nabla}_k \pi^{ki}$, and this leads to
the following equation
\begin{eqnarray}
\dot{\Gamma} = \frac{\xi - 3\dot{\alpha}}{\alpha} \Gamma -
\frac{1}{4} \kappa \bar{\rho} \overline{u^2}  + 3
\overline{\Theta^2}. \label{gammadot2}
\end{eqnarray}
Comparison of (\ref{gammadot2}) to (\ref{gammadot}) shows that for
this regime
\begin{eqnarray}
\psi =  \frac{3}{2} \bar{\rho} \overline{u^2}.
\end{eqnarray}

\medskip

{\em b. Perturbations not affecting the extrinsic curvature scalar.}
This regime includes the case of transverse-traceless tensor
perturbations and transverse vector perturbation for both variables.
In these cases, $\delta (\pi/\sqrt{h})= 0$ and hence,
 $\psi = 0 $. Then,
\begin{eqnarray}
\dot{\Gamma} =  \frac{\xi - 3 \dot{\alpha}}{\alpha} \Gamma - \kappa
\bar{\rho} \overline{u^2}+    3 \overline{\Theta^2}
\label{gammadot3}
\end{eqnarray}

\medskip

{\em c. Vector perturbations.} In this regime, the terms
corresponding to $\lambda^i$ and $\bar{\nabla}_k \delta \pi^{ki}$
dominate in the fluctuations of the metric and momentum variables
respectively. A two-tensor $A^{ij}$ corresponds to vector
perturbations if it can be written in the form
\begin{eqnarray}
A^{ij}(x) = \sum_{\bf q} \left(\frac{q^i \beta^j}{q^2} + \frac{q^j
\beta^i}{q^2} - (\beta \cdot q) \frac{q^i q^j}{q^4}\right)
\alpha_q(x),
\end{eqnarray}
where $q$ denotes the wave-vectors, corresponding to eigenvalues of
the Laplace equation for the metric ${}^0h_{ij}$ and $\beta^i$ a
vector-valued function of $q$. Assuming that $\delta \pi^{ij}$ and
$\delta (\sqrt{h} h^{ij})$ are of this form, we find
\begin{eqnarray}
\delta(\pi/\sqrt{h}) = -
\frac{\bar{\nabla}_i}{\bar{\nabla}^2}\left(\frac{\bar{\rho} u^i}{2}
+\frac{8 \xi }{\kappa \alpha} \lambda^i \right).
\end{eqnarray}
It then follows that
\begin{eqnarray}
\psi = \langle u^i {\bf P}_{ij}\frac{\bar{\rho} u^j}{2} \rangle  ,
\label{psi2}
\end{eqnarray}

where ${\bf P}_{ij} = \frac{\bar{\nabla}_i
\bar{\nabla}_j}{\bar{\nabla}^2}$ is the projector to the
longitudinal part of the vector fluctuations.

If the vector perturbations are purely longitudinal, then
\begin{eqnarray}
\psi =  \frac{\bar{\rho}}{2} \overline{u^2}.
\end{eqnarray}
At the other end, if the vector perturbations are purely transverse,
then we fall back to case b) and $\psi = 0 $.

\medskip

{\em d. Mixed perturbations.} In the previous three cases, the
perturbations of the metric $\delta h_{ij}$ and of the momentum
$\delta \pi^{ij}$ were of the same type. In principle, other cases
are possible. For example, the perturbations
 $\delta \pi^{ij}$ may be purely conformal and the perturbations
 $\delta h_{ij}$ longitudinal. This is an acceptable  region of the
 gravitational state space, since the only restriction is that the
 metric $h_{ij}$ and the momentum $\pi^{ij}$ satisfy the Hamiltonian
 constraint. For small perturbations, this regime is not
 dynamically stable (i.e., the system does not spend much time in
 the corresponding
 phase space region), because the linearized
 equations of motion couple perturbations of the same type.
 However, this regime might contain be stabilized
 when large perturbations are considered.

 An interesting case is that of
 traceless  perturbations $\delta \pi^{ij}$ and longitudinal
 perturbations $\delta h_{ij}$, for which

 \begin{eqnarray}
\psi = \frac{2 \xi}{\kappa \alpha} \langle u^i {\bf P}_{ij}
\lambda^j \rangle. \label{asymm}
 \end{eqnarray}

\hspace{1cm}

\paragraph{Evolution of $\overline{\delta R}$.}

The scalar curvature $R$ is a function on the state space, and under
Hamilton's equations evolves as
\begin{eqnarray}
\dot{R} = \frac{2N}{\sqrt{h}} R_{ij} (\pi^{ij} - \frac{1}{2} \pi
h^{ij}) + u^i \nabla_i R.
\end{eqnarray}

To second-order in the perturbation and ignoring the ultra-local
terms, we find
\begin{eqnarray}
\dot{\overline{\delta R}} = - \frac{2 \xi}{\kappa \alpha}
\overline{\delta R}+ \frac{6 \xi}{\kappa^2 \alpha^3} \overline{u^2}
+ \langle \delta R \bar{\nabla}_i u^i \rangle + \frac{2 \xi}{\kappa
\alpha} \langle \lambda^i k_i - \kappa^i \kappa_i \rangle \nonumber
\\ + \frac{2}{\kappa} \langle \kappa_i \bar{\nabla}_j \delta \, \left(
\frac{\delta \pi^{ij} - \frac{1}{2} \pi h^{ij}}{\sqrt{h}} \right)
\rangle - \frac{2}{\kappa} \langle C^k{}_{ij} \bar{\nabla}_k \,
\delta \left( \frac{\delta \pi^{ij} - \frac{1}{2} \pi
h^{ij}}{\sqrt{h}} \right) \rangle. \label{deltaR}
\end{eqnarray}

\subsection{Autonomous backreaction equations}

In  the previous section, we derived the evolution equations for the
functions $\Gamma(t), \overline{u^2}(t)$  and $\overline{\delta
R}(t)$ that enter the backreaction equations (\ref{be1b}) and
(\ref{constrain4}) for the FRW variables $\alpha$ and $\xi$. If
these equations contained no other variables other than $\Gamma(t),
\overline{u^2}(t)$  and $\overline{\delta R}(t)$, we would obtain a
closed dynamical description of backreaction in terms of a small
number of variables. We would then have effectively substituted the
infinite-dimensional dynamical system described by the Hamilton
equations (\ref{HE1}) and (\ref{HE2}) with a finite-dimensional one.

However, the set of equations for the variables $\Gamma(t),
\overline{u^2}(t)$  and $\overline{\delta R}(t)$ is not autonomous.
It depends on histories of additional variables, and these histories
can be determined by deriving additional evolution equations. These
in turn will contain additional variables and so on. This is a
general feature in the derivation of effective equations in systems
with a large number of degrees of freedom. A   common practice in
non-equilibrium statistical mechanics is to enforce  closure on a
system of effective equations for relevant variables by considering
specific regimes,  making simplifying assumptions, or introducing
phenomenological parameters.

In ordinary statistical mechanical systems, simplifications occur
because we have some knowledge for the system that allows reasonable
estimations about the state of the non-relevant variables, like, for
example, the assumption of local equilibrium or of molecular chaos
in the treatment of gases. However, in cosmology our present
knowledge of the statistical behavior of the inhomogeneities is not
detailed enough for this purpose.

A simplification that allows for the derivation of a set of closed
backreaction equations is to assume that the term involving the
perturbation $\overline{\delta R}$ of the scalar curvature in Eq.
(\ref{constrain4} is negligible in comparison to the other
backreaction terms. There is no  justification for this assumption,
other than the simplification it entails to  the backreaction
equations; in particular, it does not follow from any observed
properties of cosmological perturbations.
 However, it suffices to require that the contribution of
$\overline{\delta R}$ to the effective equations is much smaller
than {\em only one} of the other backreaction terms. This terms
dominates over the others as it does over $\overline{\delta R}$. It
is not necessary to specify which one of these terms is dominant: we
keep them all in the solution of the equations.  We must remember
this restriction when the issue of the validity of the
approximations involved is to be considered.\footnote{ In fact, it
is possible to find  regimes leading to a closed set of backreaction
equation without assuming that $\overline{\delta R}$ vanishes, but
these regimes involve many additional conditions with no clear
physical or geometric meaning.}

Setting $\overline{\delta R} \simeq 0$, the backreaction equations
under consideration are

\begin{eqnarray}
\dot{\alpha} = \xi (1 + \frac{1}{2} \overline{u^2}) - \frac{1}{3}
\Gamma \alpha \label{be1c} \\
\xi^2 = \frac{\kappa \mu_0}{6 \alpha} (1 + \frac{1}{2}
\overline{u^2}) - 1 \\ \label{be2c} \dot{\Gamma} = \frac{\xi -
3\dot{\alpha}}{\alpha} \Gamma - \frac{ \kappa \mu_0}{4 \alpha^3}
\overline{u^2}  + \frac{\kappa}{2} \psi+ 3 \overline{\Theta^2},
\label{be3c}
\end{eqnarray}
where $\overline{u^2}$ and $\overline{\Theta^2}$ are functions of
time given by Eqs. (\ref{u2t}) and (\ref{p2t}) respectively.

This set of equations can be made closed only if $\psi$ can be
written as a function of the other variables. We showed in section
5.1 that this is possible for the specific cases a) and b). We shall
also show that there are physically reasonable approximations that
allow closure also for the case of vector fluctuations.

For vector perturbations it is reasonable to consider an
interpolation between the purely longitudinal and the purely
transverse cases. The simplest interpolation involves a simple
parameter $\epsilon \in [0, 1]$, which takes the value $\epsilon =
0$ for transverse vector perturbations and the value $\epsilon = 1 $
for purely longitudinal perturbations. The parameter $\epsilon$ is
introduced by postulating an approximation
\begin{eqnarray}
\langle u^i {\bf P}_{ij}\frac{\bar{\rho} u^j}{2} \rangle \simeq
\epsilon \langle u^i \bar{h}_{ij}\frac{\bar{\rho} u^j}{2}  \rangle,
\label{approx8}
\end{eqnarray}
from which we obtain
\begin{eqnarray}
\psi =  \frac{\epsilon}{2} \bar{\rho} \overline{u^2}.
\end{eqnarray}

In order to  understand the meaning of the approximation
(\ref{approx8}) and to provide an interpretation for the parameter
$\epsilon$, we next consider a simple model for the perturbations.

\subsubsection{A model for  perturbations.} Let us assume that
perturbations $\delta h_{ij}(x)$ and $\delta \pi^{ij}(x)$ are
concentrated in $N$ non-overlapping regions $U_n$ of compact support
on $Z$,  and that they vanish elsewhere. This implies that $u^i(x) =
\sum_n u^i_{(n)}(x)$, where $u^i_{(n)}(x)$ is a field that has
support only in the region $U_n$. We also assume that the regions
$U_n$ are much smaller than the curvature radius of the homogeneous
metric $\bar{h}_{ij}$, so that
 they are adequately described by  Cartesian coordinate systems corresponding to the geodesics of
$\bar{h}_{ij}$. We denote by $V_n$ the volume of each region $U_n$
with respect to the metric $\bar{h}_{ij}$   and by $v_n = V_n/V$ the
corresponding relative volume ($V$ is the volume of $\Sigma$).

\paragraph{`Rigid' perturbations.} The simplest approximation
follows from the assumption that the regions $U_n$ are `rigid', in
the sense that they can be accurately described by the spatial
averages $\bar{u}^i_{(n)}$  of the fields $u^i_{(n)}(x)$. (The
spatial average is taken with respect to the local Cartesian
coordinate system.) This means that $u^i_{(n)}(x) \simeq
\bar{u}^i_{(n)} \chi_{U_n}(x)$, where $\chi_{U_n}$ is a smeared
characteristic function of $U_n$. \footnote{ A smeared
characteristic function corresponds to a boundary that is not
sharply defined. It is convenient in order to avoid problems arising
from differentiation.} We note that  this case corresponds to
$\Theta_{ij} \simeq 0$.

In this approximation
\begin{eqnarray}
\psi = \frac{1}{N} \sum_{n=1}^{N} v_n {}^L\bar{u}^i_{(n)}
\frac{\bar{\rho} \bar{u}^i_{(n)}}{2} , \label{psisum}
\end{eqnarray}

where
\begin{eqnarray}
{}^L\bar{u}^i_{(n)} = \frac{1}{V_n} \int d^3 x {\bf P}^i{}_j
u^j_{(n)}(x) \chi_{U_n}(x) = \frac{1}{V_n} \int \frac{d^3{\bf k}}{(2
\pi)^3} \frac{k^i k_j}{k^2} |\tilde{\chi}_{U_n}({\bf k})|^2,
\label{ulb}
\end{eqnarray}

where $\tilde{\chi}_{U_n}({\bf k})$ is the Fourier transform of
$\chi_U$.

The value of ${}^L\bar{u}^i_{(n)}$  depends on the shape of the
region $U_n$. For a region with no preferred direction (i.e., a
ball) $\tilde{\chi}_{U_n}({\bf k})$ is a function only of ${\bf
k}^2$ and Eq. (\ref{ulb}) yields
\begin{eqnarray}
{}^L\bar{u}^i_{(n)} = \frac{1}{3} \bar{u}^i_{(n)}.
\end{eqnarray}

Then Eq. (\ref{psisum}) leads to Eq. (\ref{approx8}) with $\epsilon
= 1/3$.

A second interesting case is that of fluctuation regions $U_n$ with
one  preferred direction. Let $r^i_{(n)}$ be the unit vector
corresponding to this direction;  $\tilde{\chi}_{U_n}$ is a function
of ${\bf r}_{(n)} \cdot{\bf  k}$ and ${\bf k}^2$. Eq. (\ref{ulb})
yields
\begin{eqnarray}
{}^L\bar{u}^i_{(n)} = \frac{1 - s_n}{2} \bar{u}_{i(n)} + \frac{3 s_n
-1}{2} (\bar{u}^k_{(n)} r_{k(n)}) n_{i(n)},
\end{eqnarray}
where  $s_n(x) \in [0, 1]$ is defined as
\begin{eqnarray}
s_n = \frac{1}{V_n} \int \frac{d^3{\bf k}}{(2 \pi)^3}  \frac{{\bf
r}_{(n)} \cdot{\bf k}}{{\bf k}^2} |\tilde{\chi}_{U_n}({\bf r}_{(n)}
\cdot{\bf k}, {\bf k}^2)|^2
\end{eqnarray}

In general, $s_n$ depends on the relative size of the preferred
direction: it takes values close to 1 for disk-shaped regions and
close to 0 for rod-shaped ones.

 Different assumptions about the distribution of the directions of
$r^i_{(n)}$ corresponds to different values of $\epsilon$. For
example, we may assume that  the directions $r^i_{(n)}$ and the
shape parameters $s_n$ are not statistically correlated (i) with
each other, (ii) with
 $\bar{u}^i_{(n)}$, and (iii)along
 different regions $U_n$.
Then, Eq. (\ref{psisum}) yields (\ref{approx8}) with $\epsilon =
\frac{1}{3}$. Hence, the shape of the inhomogeneity regions does not
affect the value of $\epsilon$ if its determining parameters can be
treated as uncorrelated random variables.

In presence of correlations, the value of $\epsilon$ differs.  For
example, if the mean velocity of the inhomogeneous regions tends to
align along the axis $r^i_{(n)}$, then ${}^L\bar{u}^i_{(n)} = s_n
\bar{u}^i_{(n)}$. Substituting into Eq. (\ref{psisum}) (and assuming
no correlations for the value of $s_n$ in different regions), we
find $\epsilon = \frac{1}{N} \sum_n \bar{s}_n$, i.e., $\epsilon$ can
take any value in the interval $[0, 1]$, depending on the
distribution of shapes for the inhomogeneity regions. Similarly, if
$u^i_{(n)}$ tends to lie on the plane normal to $r^i_{(n)}$, we find
that $\epsilon \simeq 1 - \frac{1}{N} \sum_n \bar{s}_n$.

\paragraph{ Expanding (contracting) perturbations.} If the mean
velocities $\bar{u}^i_{(n)}$ of the regions $U_n$ are small compared
to the values of $u^i_{(n)}(x)$, then the regions $U_n$ are `at
rest' in the comoving coordinates and the field $u^i_{(n)}(x)$
primarily contributes to the change of their shape. The longitudinal
part corresponds to change in volume, while the transverse part
corresponds to volume-preserving (`tidal') deformations of the
regions's shape. The parameter $\epsilon $ then corresponds to the
fraction of the (non-relativistic) kinetic energy of the fluid that
corresponds to expansion (or contraction) of the region. In this
light, the approximation (\ref{approx8}) is equivalent to the
assumption that the tidal deformations in the regions $U_n$ are not
persistent and they vanish on the mean---hence, that $\psi$ receives
contributions only from the fraction  of the velocity field that
corresponds to changes of volume in the perturbed regions.

\vspace{0.5cm}

With similar arguments, we can show that for the case of asymmetric
perturbations--Eq. (\ref{asymm}), the approximation $\langle u^i
{\bf P}_{ij} \lambda^j \rangle \simeq å \langle u^i \bar{h}_{ij}
\lambda^j \rangle$ yields
\begin{eqnarray}
\psi = \epsilon \frac{2\xi}{\kappa \alpha} \Gamma \label{psimixed}
\end{eqnarray}
\vspace{0.3cm}

The parameter $\epsilon$ is therefore introduced as a
phenomenological quantity that characterizes the state of the
perturbations (i.e. the actual three- metric) at a moment of time
$t$. In a sense, it is analogous to similar phenomenological
quantities that are introduced in non-equilibrium thermodynamics,
e.g., coefficients of viscosity, heat transfer and so on. The
approximation (\ref{approx8}) is not expected to be meaningful over
the whole of the gravitational state space, but only in a specific
region, corresponding, to the cases considered above. $\epsilon$ is
expected to change in time, as the perturbations evolve. One then
needs also assume that such changes occur in a time-scale much
larger than the ones relevant to the resulting effective equations.

Either with the introduction of the $\epsilon$ parameter, or by
restricting to the specific regimes considered earlier, Eqs.
(\ref{be1c}-\ref{be3c}) together with Eqs. (\ref{u2t}) and
(\ref{p2t}) provide a consistent and closed set of backreaction
equations to leading order in perturbations around the symmetry
surface. These equations hold if the perturbation terms are
significantly smaller to the ones of  FRW evolution. Applied to
cosmology, such backreaction terms could perhaps lead to corrections
on the order of a few percent from the FRW evolution.

\subsection{Solutions}

We next solve the backreaction equations (\ref{be1c}-\ref{be3c}) for
the scale factor, assuming that the perturbation parameters $\Gamma,
\overline{u^2}$ and $\overline{\Theta^2}$ are small so that only
their first-order contributions to their evolution equations are
signifcant. We assume that the geometry is approximately flat so
that in Eq. (\ref{be2c}) $\xi >>1$ and $\sqrt{\frac{\kappa
\mu_0}{\alpha}} >> 1$; this assumption allows for an analytic
solution of the differential equations.

Using Eqs. (\ref{u2t}), (\ref{p2t}) and (\ref{be3c}) we find
\begin{eqnarray}
\dot{\Gamma} =  -2 \frac{\dot{\alpha}}{\alpha} \Gamma - \frac{f
\kappa \mu_0 u^2_0 \alpha_0^2}{\alpha^5} + 3 \frac{\theta_0^2
\alpha_0^4}{\alpha^4},
\end{eqnarray}
where we set $\alpha(t_0) = \alpha_0$, $\overline{u^2}(t_0) = u^2_0$
and $\overline{\Theta^2}(t_0) = \theta_0^2$, for some time $t_0$.
The parameter $f$ takes the values $\frac{1}{4}$, $1$ and $1 -
\frac{\epsilon}{4}$ for the cases a, b and c of Sec. 5.2,
respectively.

Using Eq. (\ref{be1c}) and keeping terms of leading order to the
perturbations we obtain
\begin{eqnarray}
\frac{d \Gamma}{d \alpha} = -\frac{2}{\alpha} \Gamma - \frac{\sqrt{6
\kappa \mu_0} f u_0^2 \alpha_0^2}{a^{9/2}} + \frac{3 \theta_0^2
\alpha_0^4\sqrt{\frac{6}{\kappa \mu_0}}}{\alpha^{7/2}}
\end{eqnarray}

This is a linear inhomogeneous equation with solution
\begin{eqnarray}
\Gamma = \left(\frac{\alpha_0}{\alpha}\right)^{2} \left\{ \Gamma_0 +
b_1 \left[ \left(\frac{\alpha_0}{\alpha}\right)^{3/2} - 1 \right] -
b_2 \left[ \left(\frac{\alpha_0}{\alpha}\right)^{1/2} - 1
\right]\right\}, \label{gammaalpha}
\end{eqnarray}
where $b_1 = \sqrt{\frac{6 \kappa \mu_0}{\alpha_0^3}} \frac{2f
u^2_0}{3}$ and $b_2 = 6 \theta_0^2 \sqrt{\frac{6 \alpha_0^3}{\kappa
\mu_0}} $, and $\Gamma_0 = \Gamma(t_0)$.

Substituting into Eq. (\ref{be1c}) and changing into the variable $u
= (\alpha/\alpha_0)^{3/2}$, we find
\begin{eqnarray}
\frac{2}{3} \dot{u} = \sqrt{\frac{\kappa \mu_0}{6}} \left[ 1 +A
u^{-4/3} + Bu^{-2/3} + C u^{-1/3} \right],
\end{eqnarray}
in terms of the constants
\begin{eqnarray}
A &=& \frac{u^2_0}{\alpha_0^{3/2}} \left(\frac{3}{4} - \frac{4f}{3} \right)\\
B &=& 12 \frac{\theta_0^2
\alpha_0^{3/2}}{\kappa \mu_0} \\
C &=& - \sqrt{\frac{2}{3 \kappa \mu_0}} \Gamma_0 + \frac{4u_0^2}{3
\alpha_0^{3/2}} - 12 \frac{\theta_0^2 \alpha_0^{3/2} }{\kappa
\mu_0}.
\end{eqnarray}
We then obtain the solution to the backreaction equations
\begin{eqnarray}
\int_1^{\left(\frac{\alpha}{\alpha_0}\right)^{3/2}}  \frac{du}{ 1 +A
u^{-4/3} + B u^{-2/3} + C u^{ -1/3}} = \frac{3}{2}
\sqrt{\frac{\kappa \mu_0}{6}} (t -t_0). \label{solution}
\end{eqnarray}

\begin{figure}[tbp]
\includegraphics[height=5cm]{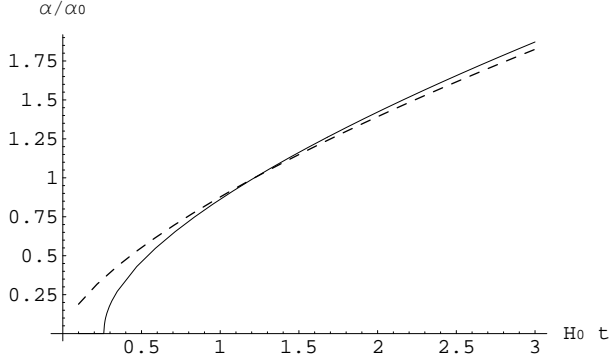} \caption{ \small The normalised scale factor $\alpha/\alpha_0$  as a function
of time $t$ in units of the Hubble time $H_0^{-1}$; $H_0 =
\frac{\dot{\alpha}}{\alpha}(t_0)$, for $\theta_0^2 = 0.01, u_0^2 =
0.0001, \Gamma_0 = -0.1$. The dashed line shows the FRW solution.}
\end{figure}

In Fig. 1 a plot is shown of a solution to Eq. (\ref{solution}) in
comparison to the FRW-solution. The two curves diverge at early
times. However, in this regime the backreaction parameters are
large; for examples $\overline{\Theta^2}$ grows at small $\alpha$ as
$\sim \alpha^{-4}$. The approximation employed in the derivation of
(\ref{solution}) is therefore not reliable at very early times and
the divergence may be absent in the regime of large perturbations.

The approximation is preserved for later times, and then we see that
$\alpha(t)$ essentially coincides with the FRW solution. Indeed, let
us denote by $\alpha_{FRW}(t)$ the FRW solution corresponding to
$\alpha(t_0) = \alpha_0$ ($A = B = C = 0$), and by $\alpha(t)$ the
solution corresponding to $\alpha(t_0) = \alpha_0$ for non-zero but
small values of the parameters $A, B$ and $C$. The difference
$\delta \alpha (t) = \alpha(t) - \alpha_{FRW}(t)$ has the following
asymptitic behavior
\begin{eqnarray}
|\delta \alpha(t)| \sim C \alpha(t).
\end{eqnarray}
This implies that the change in the Hubble parameter $H(t) =
(\dot{\alpha}/\alpha)(t)$ is $|\delta H (t)| \sim C$, i.e., the FRW
solution is stable with respect to small perturbations.

This result holds for symmetric perturbations (e.g., cases a, b and
c of Sec. 5.2). If, however, the perturbations are mixed (case d of
Sec. 5.2), the conclusions may be different. The solution to the
backreaction equations for $\psi$ given by Eq. (\ref{psimixed}) can
be obtained through a similar procedure to the one leading to Eq.
(\ref{solution}):

\begin{eqnarray}
\int_1^{\left(\frac{\alpha}{\alpha_0}\right)^{3/2}}  \frac{du}{ 1
+A' u^{-4/3} + B' u^{-2/3} + C' u^{\frac{2 \epsilon -1}{3}}} =
\frac{3}{2} \sqrt{\frac{\kappa \mu_0}{6}} (t -t_0),
\label{solution0}
\end{eqnarray}
in terms of constants $A', B'$ and $C'$ that depend on the values of
$\theta_0, u_0$ and $\Gamma_0$. In this case, the deviation $\delta
\alpha$ from the FRW evolution behaves asymptotically as
\begin{eqnarray}
|\delta \alpha(t)| \sim C' [\alpha(t)]^{1+\epsilon}.
\end{eqnarray}

Hence, for large values of $t$, the change in the Hubble parameter
grows as $\delta H(t) \sim [\alpha(t)]^{\epsilon}$. Even a small
value of $\epsilon$ produces an evolution that diverges from the FRW
prediction. However, an assumption in this derivation is that the
value of $\epsilon$ at time $t_0$ remains constant in time, which,
in the linearized treatment of perturbations, is only plausible for
very special conditions at $t = t_0$. Nonetheless, this result
strongly suggests that even small perturbations may make problematic
the extrapolation of the FRW evolution into the asymptotic future.

\section{Large perturbations}

\subsection{The approximation scheme}
Part of the motivation for this work is the idea that the  cosmic
acceleration, deduced from the supernova data, can be accounted by
backreaction effects rather than dark energy. This is only possible
if  the perturbations are large: the corresponding terms must be of
the same order of magnitude as the ones of  FRW evolution. The
approximation used in the previous section is therefore not
sufficient for this case: terms of third and higher order to the
perturbations $\delta h_{ij}, \delta \pi^{ij}$ from the symmetry
surface $\Gamma_0$ are expected to become significant.

The effective equations describing the backreaction of large
perturbations depend strongly on the region of the gravitational
phase space to which they correspond. Inevitably, in order to
construct such a set of equations for the cosmological setting, it
is necessary to have substantial information about the nature and
behavior of inhomogeneities in the present-day Universe. In absence
of such information, any model describing backreaction of large
perturbations must proceed by more or less {\em ad hoc} assumptions
about the nature of inhomogeneities. Such assumptions are necessary
both for the implementation of a consistent approximation scheme
(i.e., which effective variables can be thought of as "large"?) and
for the closure of the system of effective equations.

 In this section, we will explore one regime of large
perturbations, which corresponds to a rather natural generalization
of the perturbation scheme developed in the previous section.  We do
not construct a closed set of backreaction equations for the general
case as in the previous section; the problem is substantially more
involved. Our study is mainly kinematical: we explore the
plausibility of the idea that large perturbations around an FRW
universe  can be held responsible for the observed acceleration of
the scale factor.

To simplify our calculations we assume that the deviation velocities
are very small, namely that $\overline{u^2} << 1$. Indeed, this
condition is born out by observations, if we identify $u_i$ with
velocities of galaxy clusters---see \cite{Kash} and references
therein. The term $\overline{u^2}$ cannot be, therefore, responsible
for the substantial divergence from the FRW prediction that would be
necessary, in order to account cosmic acceleration in terms of
backreaction effects. We will  drop such terms from our
calculations.
 As in the previous section, we shall also ignore the
the perturbations of the curvature scalar and consider almost flat
three-geometries. Consequently,  the only large backreaction
variable in the effective equations (\ref{be1b}--\ref{constrain4})
is $\Gamma$.

In order to construct backreaction equations for large
perturbations,  we  employ the following approximation scheme.
 We  assume that the perturbations
$\delta h_{ij}$ and $\delta \pi^{ij}$ are small in the ultra-local
sense, but that they may be large when acted upon by derivatives.
This  means that the backreaction terms may become large because of
the strong spatial variation of the inhomogeneities. In particular,
we assume that the perturbations $\delta h_{ij}$ and $\delta
\pi^{ij}$ are a fraction of order $\epsilon$ of the averaged
variables $\bar{h}_{ij}$ and $\pi^{ij}$ (with respect to a matrix
norm), and that the variation of the perturbations takes place at a
scale $l$ on the spatial surface. Expanding to the $n-th$ order of
perturbations around the symmetry surface, a term that involves $m$
derivatives will be of order $\epsilon^n/l^m$. The approximation
scheme we shall use in this section involves keeping all terms such
that $m - n \geq 0$, which means that $\epsilon$ is of the same
order of magnitude with $l$.

\subsection{The effective equations}

In the approximation scheme described above, the backreaction
equations (\ref{be1b}--\ref{constrain4}) remain unchanged. The
higher order backreaction terms are characterized by negative values
of $m-n$. Again, as in the previous section, one of these equations
can be used to determine the functional dependence of the curvature
perturbations $\overline{\delta R}$ in terms of the other
parameters. Dropping the contribution of the $\overline{u^2}$ terms,
Eqs. (\ref{be1b}--\ref{constrain4}) take a simple form

\begin{eqnarray}
\dot{\alpha} = \xi  - \frac{1}{3} \Gamma \alpha \label{be1e}
\\
 \xi^2 + 1 =    \frac{\kappa \mu_0}{6
\alpha}  \label{be2e}
\end{eqnarray}

 Eq. (\ref{ldot1}) for the evolution of $\lambda^i$ is valid
also within this scheme. Together with (\ref{ui}), it leads to the
following equation for $\Gamma$

\begin{eqnarray}
\dot{\Gamma} =  \frac{\xi - 3 \dot{\alpha}}{\alpha} \Gamma + 3
\overline{\Theta^2} + \omega \label{gammadot4}
\end{eqnarray}

Compared to Eq. (\ref{gammadot}), terms $\overline{u^2}$ have been
dropped (in particular $\psi \simeq 0$) and there appears a new
term.
\begin{eqnarray}
\omega = \langle \bar{\nabla}_i \lambda^i u_k u^k \rangle + 2
\langle \bar{\nabla}_k u_i C^{lik}u_l \rangle - \langle
\bar{\nabla}_i u^i \lambda^k u_k \rangle. \label{omega}
\end{eqnarray}

In the derivation of Eq. (\ref{gammadot4}), we dropped ultralocal
terms $\delta h_{ij}$ when they are additive to terms
$\bar{h}_{ij}$: for example, we approximated $\langle \bar{\nabla}_i
\lambda^i u_k u_l h^{kl}\rangle \simeq \langle \bar{\nabla}_i
\lambda^i u_k u_l \bar{h}^{kl}\rangle$.

To simplify  Eq. (\ref{omega}), we further assume that the tensor
$\Theta_{ij} = \frac{1}{2} (\bar{\nabla}_i u_j + \bar{\nabla}_j
u_i)$ is isotropic, i.e., that
\begin{eqnarray}
\Theta_{ij} = \Theta \bar{h}_{ij}, \label{thdiag}
\end{eqnarray}
where $\Theta$ is a scalar function on $Z$. This means that
expansion dominates over shear in the congruence of comoving
observers.

Condition (\ref{thdiag}) implies that

\begin{eqnarray}
\omega = -3 \langle \Theta \lambda^i u_i \rangle. \label{omega2}
\end{eqnarray}

The right-hand-side of (\ref{omega2}) corresponds to a spatial
integral over $Z$.  Schwarz' inequality then applies
\begin{eqnarray}
|\omega|^2 \leq 9 \langle \Theta^2 \rangle \left(\langle (\lambda^k
u_k )^2 \rangle \right). \label{omega3}
\end{eqnarray}

We define the deviation $\delta \Gamma$ for $\lambda^k u_k$ as
\begin{eqnarray}
(\delta \Gamma)^2 = \langle (\lambda^ku_k)^2 \rangle - \Gamma^2,
\end{eqnarray}

Eq. (\ref{omega3}) then can be written as
\begin{eqnarray}
|\omega| \leq 3 \sqrt{\overline{\Theta^2}} \sqrt{\Gamma^2 + (\delta
\Gamma)^2)}.
\end{eqnarray}

 This implies that
\begin{eqnarray}
\omega =  3 \eta \sqrt{\overline{\Theta^2}} \Gamma \sqrt{1 +
\left(\delta \Gamma / \Gamma \right)^2}, \label{omega4}
\end{eqnarray}

for $|\eta| < 1$. The parameter $\eta$ is, in general, a function of
time. We may view it as a phenomenological parameter characterizing
the present state of fluctuations. It characterizes the strength of
correlations between the backreaction parameters
$\overline{\Theta^2}$ and $\Gamma$. If $\eta = 0$, then there is no
correlation and $\omega = \langle \Theta \rangle \langle \lambda^k
u_k \rangle = 0$, since $\langle \Theta \rangle  = \frac{1}{3}
\langle \bar{\nabla}_ku^k \rangle = 0$. If $\eta = \pm 1$ then the
Schwarz inequality (\ref{omega3}) is saturated and $\Theta = a
\lambda^k u_k$ for some constant $a$ at all points of $Z$.

\subsection{Cosmological models with backreaction}

With the assumptions state above, the set of equations for the
backreaction consists of Eqs. (\ref{be1e}, \ref{be2e},
\ref{gammadot4}, \ref{omega4}) and the expressions for $\psi$
provided in the previous section. In general, this does not
constitute a closed set of equations, because we have not provided
an evolution equation for $\overline{\Theta^2}$ that appears in Eq.
(\ref{omega4}). The set of equations only closes for the case of
longitudinal `rigid' perturbations considered in section 5.3.
 However, these expressions suffice for establishing the plausibility that
cosmic acceleration is possible in the regime of large perturbations
we have considered here.

We  assume that the effective FRW spacetime is almost flat, i.e.
that Eq. (\ref{be2e}) implies that
\begin{eqnarray}
\xi = \sqrt{ \frac{\kappa \mu_0}{6 \alpha}}.
\end{eqnarray}
The following equations then apply
\begin{eqnarray}
 \left( \frac{\dot{\alpha}}{\alpha} + \frac{1}{3} \Gamma
\right)^2 &=&
\frac{\kappa \bar{\rho}}{6} \label{constr+}\\
 \frac{\ddot{\alpha}}{\alpha} &=& - \frac{\kappa \bar{\rho}}{12} -
 \frac{1}{6} \Gamma \frac{\dot{\alpha}}{\alpha} +
 \frac{1}{18}\Gamma^2 - \frac{1}{3} \dot{\Gamma}, \label{accel} \\
\dot{\Gamma} &=& - 2 \Gamma \frac{\dot{\alpha}}{\alpha} +
\frac{1}{3}
 \Gamma^2 + 3 \left(\overline{\Theta^2} + \eta
\sqrt{\overline{\Theta^2}} \Gamma \sqrt{1 + \left(\delta \Gamma /
\Gamma \right)^2 }\right), \label{gamma8}
\end{eqnarray}

Substituting (\ref{gamma8}) into (\ref{accel} we obtain an equation
for the acceleration $\ddot{\alpha}$:
\begin{eqnarray}
\frac{\ddot{\alpha}}{\alpha} = - \frac{\kappa \bar{\rho}}{12} +
\frac{1}{2} \Gamma \frac{\dot{\alpha}}{\alpha} - \frac{1}{18}
\Gamma^2 -  \left[\overline{\Theta^2} + \eta
\sqrt{\overline{\Theta^2}} \Gamma \sqrt{1 +\left( \delta \Gamma /
\Gamma \right)^2 } \right] \label{accel2}
\end{eqnarray}

Comparing Eqs. (\ref{constr+}--\ref{accel2}) to the standard FRW
equations, we note the following. First, the constraint equation
(\ref{constr+}) is dissimilar to the FRW equations $H^2 =
\frac{\kappa}{6} \rho$ irrespective of the theory's matter content.
Dark energy or a cosmological constant or any other form of matter
will be added to the energy density term. The backreaction term
$\Gamma$ is additive to the Hubble parameter $H$, which appears
squared in the FRW equations. A negative-valued $\Gamma$ implies
that the actual value of the Hubble parameter is larger than the one
predicted from the matter content of the FRW solution.

 Moreover, the presence of the $\Gamma$-dependent terms does not allow us to bring
 Eq. (\ref{accel2}) into an FRW-like form
 \begin{eqnarray}
\frac{\ddot{\alpha}}{\alpha} = - \frac{\kappa}{12} (\bar{\rho} + 3
P_{eff})
 \end{eqnarray}
in terms of some "effective pressure" some $P_{eff}$. Again, this
indicates how the backreaction terms differ in their structure to
the ones that can be obtained from any form of matter. The presence
of $\Theta$-dependent terms in the right-hand of
 Eq. (\ref{accel2}) would suggest that at least this part of the acceleration could be
 interpreted as some sort of bulk viscosity pressure generated by the
 "gravitational interaction" between the inhomogeneities--see \cite{bf} for a possible relation of bulk viscocity to
 cosmic acceleration. However, the dependence of the acceleration on
 pressure is quadratic on $\Theta$, instead of linear as would be
 expected from an ordinary viscous fluid. Overall, the structure of
 the backreaction terms are very different from what one would
 expect from any form of matter.

We next consider Eq. (\ref{constr+}) at the present moment of time
$t = t_0$. Defining the Hubble constant $H_0 =
\left(\dot{\alpha}/\alpha \right) (t_0)$ and $\Omega_m = \frac{
\kappa \bar{\rho}(t_0)}{6 H_0^2}$ we obtain
\begin{eqnarray}
1 + \frac{\Gamma(t_0)}{3 H_0} = \sqrt{\Omega_m}.
\end{eqnarray}

If $\Omega_m < 1$, then $\Gamma(t_0) < 0$. We introduce the
parameter $\gamma_0 := - \Gamma(t_0)/(3 H_0)$; substituting into
(\ref{accel}) we calculate  the deceleration parameter $q_0 = -
\frac{\ddot{\alpha}{\alpha}}{\dot{\alpha}^2}(t_0)$ as

\begin{eqnarray}
q_0 = \frac{1}{2} + \frac{1}{2} \gamma_0 + \gamma_0^2 + 3 (
\theta_0^2 -  \eta \theta_0 \gamma_0 \sqrt{1 + (\delta \gamma_0
/\gamma_0)^2}), \label{q0}
\end{eqnarray}
where $\theta_0 = \sqrt{\overline{\Theta^2}}(t_0)$, and we also
defined $\delta \gamma_0 := - (\delta \Gamma)(t_0)/(3H_0)$.

\subsection{An exactly solvable regime}

Before proceeding to an examination of Eq. (\ref{q0}), we note that
in the regime of "rigid" perturbations described in Sec. (5.2), the
system of backreaction equations is closed. In this case,
$\overline{\Theta^2} = 0$, and the set of backreaction equations
(\ref{be1e}, \ref{be2e}, \ref{gammadot4}) reduces to
\begin{eqnarray}
\dot{\alpha} = \sqrt{\frac{\kappa \mu_0}{6 \alpha}} - \frac{1}{3}
\Gamma \alpha \label{be6} \\
\dot{\Gamma} = - 2  \frac{\dot{\alpha}}{\alpha} \Gamma + \frac{1}{3}
\Gamma^2. \label{gammadot9}
\end{eqnarray}

This system of equations can be integrated to determine the
functional dependence between $\Gamma$ and $\alpha$.

\begin{eqnarray}
\frac{\Gamma}{H_0} \left( \frac{\alpha}{\alpha_0}\right)^{3/2} =
\sqrt{\Omega_m} + (\Gamma/\Gamma_0)^{3/2}(\alpha/\alpha_0)^{ 3}
(\Gamma_0 -  \sqrt{\Omega_m}) \label{implicit}
\end{eqnarray}

Eq. (\ref{implicit}) is an algebraic equation that can be implicitly
solved to determine $\Gamma$ as a function of $\alpha$. Substituting
$\Gamma(\alpha)$ into Eq. (\ref{be6}) then provides a closed
differential equation for the scale factor $\alpha$.

Since $\Gamma_0$ is negative, Eq. (\ref{gammadot9}) implies that it
was still negative and with larger negative value in the past. Then
Eq. (\ref{accel2}) implies that the acceleration ratio
$\ddot{\alpha} / \alpha$ was negative and with larger absolute value
in the past. It follows that Eqs. (\ref{be6}) and (\ref{gammadot9})
describe a universe that has expanded more slowly than the FRW
solution. Hence, this regime does not correspond to an evolution
compatible with the supernova data.

\subsection{Cosmic acceleration}
We  return to Eq. (\ref{q0}), and examine whether there is a regime
in which acceleration is possible ($q_0 < 0$). We note the following

\begin{itemize}
\item For $\theta_0 = 0$ acceleration is not possible.
For acceleration, it is necessary that the $\theta_0$-dependent term
is negative-valued, i.e., that $ \eta > 0 $ and that  $0 < \theta_0
<  \eta \gamma_0 \sqrt{1 +\left( \delta \gamma_0 / \gamma_0
\right)^2}$.

\item A {\em necessary} condition for acceleration ($q_0 < 0$) is that
$\eta \delta \gamma_0  > \frac{2}{3}$; this implies, in particular,
that if $\delta \gamma_0 \simeq 0$, no acceleration is possible.
\end{itemize}

\begin{figure}[tbp]
\includegraphics[height=5cm]{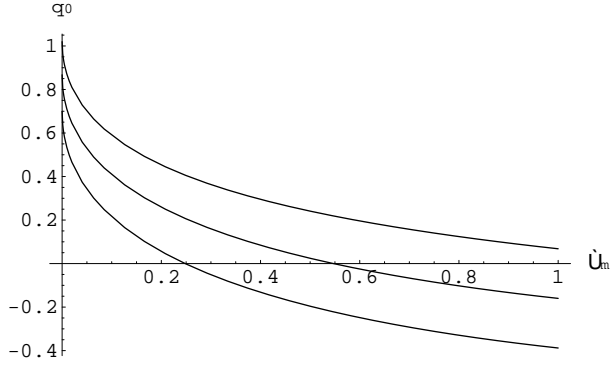} \caption{ \small Plots of the decelerations parameter $q_0$ as a function of $\Omega_m$, for
$\eta = 0.95$, $\theta_0 = 0.4$ and different values of $\delta
\gamma_0$: from top to bottom: $\delta \gamma_0 = 0.8$, $\delta
\gamma_0 = 1.0$, $\delta \gamma_0 = 1.2$. }
\end{figure}

Plots of $q_0$ as a function of $\gamma_0$ are provided in Fig. 2
for specific values of $\theta_0, \eta$ and $\delta \gamma_0$. The
region of the gravitational state space space region that gives rise
to negative $q_0$ is characterized by negative value for the
parameter $\Gamma = \langle \lambda^k u_k \rangle$. Moreover, the
field $\Theta = \frac{1}{3} \bar{\nabla}_i u^i$ must be non-zero and
positive-valued (by virtue of Eqns. (\ref{omega2}), (\ref{omega4})
and of the fact that $\eta > 0 $). This means that the expansion of
the congruence of comoving observers must be significant. In other
words there must be a strong expansion of the inhomogeneous regions,
{\em in addition to the Hubble expansion}. A substantial correlation
of the fields $\Theta$ and $\lambda^k u_k$ is also necessary.

The above are strong restrictions, but they do not require any
fine-tuning of parameters. They correspond to a fairly generic
region of the gravitational state space.

The  conclusion  above only involved a {\em kinematical} study of
 backreaction. To argue that cosmic acceleration is
a consequence of backreaction, a dynamical study is necessary.
Namely, one must demonstrate that the phase space region
corresponding to acceleration can be reached from sufficiently
generic (i.e., not fine-tuned) and physically reasonable initial
conditions. To do so one needs to write a closed set of effective
backreaction equations and study the properties of its solutions.
This will be the topic of future work.

\subsection{Special choices for the background metric}

Many studies of cosmic acceleration as arising from the
inhomogeneous nature of the universe have considered as a true
spacetime various forms of the Lemaitre-Tolman- Bondi (LTB) family
of solutions \cite{ltb}. These describe a spherically-symmetric
cosmological spacetime. In particular, void LTB models, namely,
models describing a large underdense central region surrounded by a
flat-matter dominated spacetime provide remarkable agreement with a
large fraction of cosmological date, provided we are located near
the void's center.

In principle, our method can be straightforwardly applied to
LTB-type solutions to Einstein's equations. In the canonical
formulation, such solutions correspond to three-metrics of the form
\begin{eqnarray}
ds^2 = f(\chi) d\chi^2 + g(\chi) (d\theta^2 + \sin^2 \theta d
\phi^2),
\end{eqnarray}
and momenta
\begin{eqnarray}
\pi^{ij} = \mbox{diag} [ p_f(\chi), \frac{1}{2} p_g(\chi),
\frac{1}{2} p_g(\chi)/\sin^2 \theta ] \sin \theta,
\end{eqnarray}
where $p_f, p_g$ are the conjugate momenta of the variables $f$ and
$g$ respectively, and the canonical variables satisfy the
Hamiltonian contraint in  Eq. (\ref{constr}). We assumed that the
Cauchy surfaces are three-spheres and that the perfect fluid is
dust.

The group-averaging calculus yields allows us to define the
variables $\alpha$ and $\xi$ of the effective description, through
equations $\langle h_{ij} {}^0h^{ij} \rangle = 3 \alpha^2$ and
$\langle \pi^{ij}{}^0h_{ij}/ \sqrt{{}^0h} \rangle = - 3 \kappa \xi$.
We find
\begin{eqnarray}
\alpha^2 = \frac{2}{3 \pi} \int_0^{\pi} d \chi [f(\chi) \sin^2 \chi
+ 2 g(\chi)] \\
\xi = - \frac{2}{3 \pi \kappa} \int_0^{\pi} d \chi [ p_f(\chi) +
p_g(\chi) \sin^2 \chi].
\end{eqnarray}

A study of backreaction for the LTB models would then proceed along
the lines described in sections 4 and 6. The state space of the LTB
model is infinite-dimensional, and a full study of backreaction
would be beyond the scope of this paper. However, there are some
points we can make in relation to the approximation schemes we have
employed.

In particular, we have relied on the approximation that the
ultra-local terms to the perturbations are negligible. This means
that the magnitude of the metric perturbations $\delta f = f -
\alpha^2, \delta g = g - \alpha^2 \sin^2 \chi$ must be much smaller
than $\alpha^2$ and $\alpha^2 \sin^2 \chi$ respectively. The same
holds for the perturbations to the momenta $\delta p_f$ and $\delta
p_g$. For LTB solutions that do not satisfy this condition, the
approximation scheme employed in sections 4 and 6 fails, and a
different scheme should be developed.

The derivation of Eqs. (\ref{constr+}--\ref{accel2}) employed
additional assumptions. First, a negligible value of the mean scalar
curvature $\langle R $ and of the mean-square deviation velocities
$\langle u^2 \rangle$. These conditions are restrictive, but it is
possible to find configurations that satisfy them. However, the
large degree of symmetry of the LTB solutions does not allow for the
isotropy of the tensor $\Theta_{ij}$. The deviation velocity field
$u^i$ is radial (along the $\partial/\partial \chi$ direction) and
depends only on $\chi$. As a result the only non-zero component of
$\Theta_{ij}$ is $\Theta_{\chi \chi}$. It follows that Eqs.
(\ref{constr+}--\ref{accel2}) do not hold in an LTB spacetime,
because of its high degree of symmetry.

It is sometimes suggested that the use of a spherically symmetric
spacetime as a cosmological model can be motivated by the assumption
of an implicit averaging of observations over the celestial sphere.
In light of our analysis, it is necessary to point out that such an
averaging would most probably misrepresent backreaction. An
averaging over the celestial sphere could be viewed in our formalism
as a group averaging over an action of the group $G = SO(3)$.  The
resulting distribution of deviation velocities would correspond to a
tensor $\Theta_{ij}$, whose only non-zero component would be
$\Theta_{\chi \chi}$ as above, which may be substantially different
from the "true" tensor $\Theta_{ij}$, and as such to lead to
qualitatively different behavior for backreaction. Void models, in
particular, need not refer to spherical symmetry, and it is to be
expected that a non-spherically symmetric void (especially with
respect to the distribution of deviation velocities) would lead to
different predictions from the LTB ones.

In fact, if we have reasons to believe that a family of solutions to
the Einstein equations with a high degree of symmetry are good
approximations to the "true" spacetime metric, it would be more
convenient to work directly with these solutions. The method we
developed here is intended to be used for generic "true" spacetimes:
its aim is to identify generic variables that drive backreaction,
and, subsequently, to relate them to observed quantities. In
particular, our method would be suitable for dealing with
"Swiss-cheese"-like models, which have also been studied in relation
to cosmological backreaction \cite{swiss}. In fact, the models for
the distribution of the perturbations in Sec. 5.2.1 could provide a
starting point for such a study, without the assumption of spherical
symmetry for the void regions.

Finally, we note that Eq. (\ref{constr+}) leads to a relation
between the Hubble factor $\frac{\dot{\alpha}}{{\alpha}}$ and the
quantity $z = \alpha_0/\alpha - 1$, of the form
\begin{eqnarray}
H(z) = H_0 \left[ \gamma(z) + \sqrt{\Omega_m} (1 + z)^{\frac{3}{2}}
\right], \label{hz}
\end{eqnarray}
where $\gamma(z)= - \Gamma/(3H_0)$ is a function of $z$, such that
$\gamma(0) = 1 - \sqrt{\Omega_m}$. In effect $\gamma(z)$ plays the
role of a redshift-dependent "anti-dissipation" coefficient.

In contrast, the relation between the Hubble factor and the redshift
for dark energy is of the form
\begin{eqnarray}
H_(z) = H_0 \sqrt{ (1 - \Omega_m) f(z) + \Omega_m (1 + z)^3},
\label{hure}
\end{eqnarray}
for some function $f(z)$. It is important to emphasize the
structural difference between the two expressions for $H(z)$: an
"anti-dissipation" term $\Gamma$ leads to a luminosity-distance
relation that differs strongly from the one obtained by most
reasonable energy-density terms. For this reason, it is conceivable
that good fits to the supernova data may be provided by simple
functional expressions for $\gamma(z)$ that do not correspond to
accelerated expansion.

The consideration of  perturbations of the scalar curvature would
lead to a modified expression of Eq. (\ref{hz}) of the form
\begin{eqnarray}
H(z) = H_0 \left[ \gamma(z) + \sqrt{\Omega_m (1 + z)^3 + r(z)}
\right],
\end{eqnarray}
where $r(z) = - \overline{\delta R}/(6 H_0^2 )$ incorporates the
effect of curvature perturbations. Unlike the $\gamma(z)$ term, the
curvature perturbations are additive to  the energy density for
dust; their presence in $H(z)$ is like that of a dark energy term.

However, one should be careful  before using the variable $z$ as a
measure of the observable redshift. The relation between redshift
and the scale factor should also take into account the presence of
inhomogeneities. Averaging over the inhomogeneities is expected to
introduce additional terms in the relation between the physical
redshift and the scale factor. The derivation of such a relation,
through group-averaging of a generic spacetime is the necessary next
step, before to attempt to relate the dynamical equations derived
from the present method to observable quantities. This implies that
the relation between the Hubble factor and redshift will be of the
form (\ref{hure}), where $z$ will be a functional of the physical
redshift. This would seem to imply an even stronger divergence from
the predictions of dark energy models.

\section{Conclusions}

In this paper, we developed a general procedure for the treatment of
backreaction in cosmological spacetimes. The key ingredients to the
formalism has been the averaging with respect to the isometry group
of FRW cosmologies. This allowed for the construction of a
projective map in the gravitational phase space for the gravitating
fluid, which was used in order to provide a consistent and gauge
covariant treatment of gravitational backreaction. For dust-filled
spacetimes we identified and solved the backreaction equations for
small perturbations and we identified regions of phase space, in
which accelerated expansion is possible if the perturbations are
large. A dynamical study of the case of large perturbations (aiming
to construct explicit solution of backreaction equations) will be
undertaken in a latter publication.

Some comments on the potential generalizations of the method are in
order. We have exploited here the properties of the Lagrangian
formalism of perfect fluids, in order to factor out the gauge
freedom corresponding to spatial diffeomorphisms. In principle, the
method can be applied to spacetimes with matter content other than a
perfect fluid, by taking into account the group-average of
(combinations of) the diffeomorphism constraint. There are, however,
issues related to the gauge-invariance of the perturbation expansion
around the symmetry surface that need to be explored.

Here, we described backreaction in terms of an autonomous set of
evolution equations, constructing an effective dynamical system with
a small number of variables. An alternative approach is to use
probabilistic arguments: the effective equations would then involve
a degree of stochasticity due to our ignorance of the detailed state
of the perturbations. Models, such as those of Sec. 5.2.1, may be
useful in this approach, the behavior of perturbations being encoded
in a small number of parameters characterizing an effective
probability distribution for the properties of localized
perturbations.


\begin{thebibliography}{}

\bibitem{ellis1}G. F. R. Ellis,  in General Relativity
and Gravitation (D. Reidel Publishing Co., Dordrecht),  B. Bertotti,
F. de Felice and A. Pascolini eds., 215 (1984).

\bibitem{back} S. Bildhauer,
Prog. Theor. Phys. 84, 444 (1990); S. Bildhauer and T. Futamase, T
Mon. Not. Roy. Astron. Soc. 249, 126 (1991); M. Carfora and K.
Piotrkowska,  Phys. Rev. D 52, 4393 (1995); T. Futamase, Phys. Rev.
D 53, 681 (1996); G. F. R. Ellis and W. Stoeger,  Class. Quant.
Grav. 4, 1697 (1987); H. Russ, M. H. Soffel, M. Kasai, and G.
B\"orner,  Phys. Rev. D 56, 2044 (1997); J. P. Boersma,  Phys. Rev.
D 57, 798 (1998); J. Ehlers and T. Buchert, Gen. Rel. Grav. 29, 733
(1997); G. F. R. Ellis and T. Buchert,  Phys. Lett. A.  347, 38
(2005).




\bibitem{ltb} M. N. C\'el\'erier, Astron. Astrophys. 353, 63 (200); H. Iguchi
, T. Nakamura  and K. Nakao,  Prog. Theor. Phys. 108, 809 (2002);
 H. Alnes and M.
Amarzguioui, Phys. Rev. D74 103520 (2006); H. Alnes, M. Amarzguioui
M and \O~. Gr\o n , Phys.Rev. D73, 083519 (2006); D. J. H. Chung and
A. E. Romano, Phys Rev. D74, 103507 (2006); H. Alnes and M.
Amarzguioui, Phys. Rev. D75, 023506 (2007); K. Enqvist and T.
Mattsson, JCAP 0702, 019 (2007);  T. Biswas, R. Mansouri  and A.
Notari,  JCAP 0712, 017 (2007); J. Garc\'ia-Bellido  and T. Haugb\o
elle, J. JCAP 0804, 03(2008); K. Enqvist, Gen. Rel. Grav. 40, 451
(2008);    S. Khosravi, E. Kourkchi, R. Mansouri and Y. Akrami, Gen.
Rel. Grav. 40, 1047 (2008); J. P. Zibin, A. Moss, and D. Scott,
Phys. Rev. Lett. 101, 251303 (2008).

\bibitem{de} T. Buchert, Gen. Rel. Grav. 9, 306 (2000); D. Schwarz,
astro-ph/0209584; S R\"as\"anen, JCAP02, 003 (2004); S. R\"as\"anen,
Class. Quant. Grav. 23,  1823 (2006);  E. Barausse, S. Matarrese and
A. Riotto, Phys. Rev. D71, 063537 (2005); E. W. Kolb, S. Matarrese
,A.  Notari  and A. Riotto,  Phys. Rev. D71, 023524 (2005); E. W.
Kolb, S. Matarrese and A. Riotto, New J. Phys. 8, 322 (2006); W.
God\l{}owski, J. Stelmach and M. Szyd\l{}owski, Class. Quant. Grav.
21, 3953 (2004); D. L. Wiltshire, Phys. Rev. Lett. 99, 251101
(2007); K .Bolejko, PMCPhys. A2, 1(2008); K. Bolejko and L.
Andersson, JCAP 0810, 03(2008)003; E. W. Kolb, V Marra, and S
Matarrese, arXiv: 0901.4566 (2009); D. L. Wiltshire, New J. Phys. 9,
377 (2007).

\bibitem{swiss} Brouzakis N, Tetradis N and Tzavara E,    JCAP
0702, 013 (2007;  V. Marra, E. W. Kolb, S. Matarrese, and A. Riotto,
Phys. Rev. D76, 123004 (2007); T. Biswas and A. Notari,  JCAP 0806,
021 (2008); R. A. Vanderveld,  E. E. Flanagan, and Ira Wasserman,
Phys. Rev. D78, 083511 (2008).


\bibitem{celer} M. N. C\'el\'erier, New Adv. Phys. 1, 29 (2007).

\bibitem{buchert} T. Buchert. Gen. Rel. Grav. 40, 467 (2008).

\bibitem{WI} A. Ishibashi and R. M. Wald, Class. Quantum Grav. 23,
235 (2006).

\bibitem{DGM} R. S. de Groot and P. Mazur, {\em Non-Equilibrium
Thermodynamics}, (Dover, 1984).

\bibitem{Zwan} R. Zwanzig, {\em Nonequilibrium Statistical
Mechanics}, (Oxford University Press, 2001).

\bibitem{Bal} R. Balescu, {\em Statistical Dynamics: Matter Out of
Equilibrium}, (World Scientific, 1997).

\bibitem{CaHu} E. M. Calzetta and B. L. Hu, {\em Nonequilibrium Quantum Field
Theory}, (Cambridge University Press, 2008).

\bibitem{proj} S. Nakajima, Prog. Theor. Phys. 20, 948 (1958); R.
Zwanzig, J. Chem. Phys. 33, 1338 (1960); R. Zwanzing, in {\em
Boulder Lectures on Theoretical Physics} edt. by W. E. Brittin, B.
W. Downs and J. Downs (Interscience, 1961); H. Mori, Prog. Theor.
Phys. 33, 423 (1965); B. Robertson, Phys. Rev. 144, 151 (1966).


\bibitem{pf} B. F. Schutz, Phys. Rev. D4, 3559 (1971); J. Demaret
and V. Moncrief, Phys. Rev. D21, 2785 (1980); D. Bao, J. Marsden and
R. Walton, Comm. Math. Phys. 99, 319 (1985); J. D. Brown, Class.
Quant. Grav. 10,  1579 (1993).

\bibitem{KSG} J. Kijowski, A. Smolski and A. Gornicka, Phys. Rev D41, 1875
(1990).

\bibitem{Sav} N. Savvidou, Class. Quant. Grav. 21, 615 (2004);
Class. Quant. Grav. 21, 631 (2004).

\bibitem{Kash} A. Kashlinsky1, F. Atrio-Barandela, D. Kocevski and H.
Ebeling, Ap. J. (letters) 686, L49 (2008); arXiv:0809.3733.

\bibitem{bf} J. C. Fabris, S. V. B. Goncalves, and R. de Sa'
Ribeiro, Gen. Rel. Grav. 38, 495 (2006); R. Colistete, J. C. Fabris,
J. Tossa, and W. Zimdahl, Phys. Rev. D 76, 103516 (2007).

\end{thebibliography}
\end{document}